\documentclass[preprint,showpacs,amsmath,amssymb,aps]{revtex4}

\usepackage{graphicx}
\usepackage{dcolumn}
\usepackage{bm}
\begin{document}
\title{Extension of the full-folding optical model for nucleon-nucleus 
       scattering with applications  up to 1.5 GeV} 
\author{H. F. Arellano}
\affiliation{Departamento de F\'{\i}sica,
        Facultad de Ciencias F\'{\i}sicas y Matem\'aticas \\
        Universidad de Chile, Casilla 487--3, Santiago, Chile}
\author{H. V. von Geramb}
 \affiliation{Theoretische Kernphysik, Universit\"at Hamburg \\
           Luruper Chaussee 149, D--22761, Hamburg, Germany}
\date{\today}


\begin{abstract}
The nonrelativistic 
full-folding optical model approach for nucleon-nucleus scattering is
extended into the relativistic regime.
In doing so, kinematical issues involving the
off-shell Lorentz boost of the colliding particles between the two 
nucleons and the projectile-nucleus center-of-mass reference frames
have been taken into account.
The two-body effective interaction is obtained in the framework of the
nuclear matter $g$-matrix using nucleon-nucleon optical model
potentials that fully account for the inelasticities and isobar resonances 
in the continuum at nucleon energies up to 3 GeV.
Diverse nucleon-nucleon potential models were constructed
by supplementing the basic Paris,
Nijmegen, Argonne or Gel'fand-Levitan-Marchenko inversion potentials
with complex separable terms. In each case the additional separable
terms ensured that the combination led to NN scattering phase shifts
in excellent agreement with experimental values. With each phase
shift fitting potential nuclear matter $g$-matrices have been formed
and with each of those 
relativistic full-folding optical potentials  
for nucleon-nucleus elastic scattering determined.
Application to such scattering  for projectile energies to 1.5 GeV
have been made.
Good and systematic agreement is obtained between the calculated
and measured observables, both differential and integrated quantities,
over the whole energy range of our study. 
A moderate sensitivity to off-shell effects in the differential scattering 
observables also is observed.
\end{abstract}

\pacs{03.65.Nk,   25.40.Cm,   24.10.-i}

\maketitle


 \section{Introduction}
Elastic nucleon-nucleus (NA) scattering is known now as an
excellent means of testing  nuclear structure \cite{Amo00}.
Results of the nonrelativistic theory for such scattering confidently
can be used as predictive for information required in  applied nuclear
technology, where large amounts of nuclear reaction data
--including fission cross sections at intermediate energies--
are required for various challenging applications such as 
accelerator transmutation of waste, particularly 
the elimination of long-lived radioactive wastes with a spallation source, 
accelerator-based conversion to destroy weapon grade plutonium, 
accelerator-driven energy production  to derive 
fission energy from thorium with concurrent destruction of the long-lived 
waste and without the production of nuclear weapon material, 
and accelerator production of tritium \cite{Kal96}. 
There is also a great need for such information to be the base in
analyzes of patient radiation therapy and protection.

With basic science there is the intellectual challenge 
to go beyond the physics of a single hadron and understand essential 
aspects of nuclear physics from first principles such 
as QCD \cite{Rob00}. It is generally agreed 
that the QCD Lagrangian involves  nonlinear dynamics.
Such makes it very difficult then  to understand 
nuclear physics fully from first principles, and so most
of nuclear physics phenomena are interpreted 
in terms of appropriate effective degrees of freedom. One such
view of natural phenomena in terms of energy scales ($Q$), divides  
the nuclear-hadronic scale  into the nuclear structure region 
 $Q\sim 1-10$\,MeV and the nucleon  and  nucleon-nucleon (NN)
region, with structure and substructure  scales $Q\sim 0.3-1$\,GeV.
This large separation between the hadronic energy scale and the nuclear
binding scale poses stringent difficulties to apply nonlinear QCD directly to
understand the physics of  nuclei.
However, quantitative calculations based on effective quantum field 
theory (EQF) techniques that arise from chiral symmetry provide an 
alternative approach.
At present, this method is being extended to address few- and many-nucleon 
interactions.
When combined with first principle calculations of the low energy constants
from QCD, these EQF may  provide a consistent qualitative 
understanding of properties of nuclei and of low to medium energy nuclear 
scattering.

Besides understanding the structure of nuclei from a QCD 
point of view, it is  of interest to understand the dynamical 
behavior of nucleons in the 
presence of nuclear matter. 
The relevance of modified NN scattering amplitudes in the form of
$g$-matrices in the nuclear medium with mean fields and 
Pauli blocking is well known. 
These NN amplitudes and alternative reductions 
--in the form of $t$-matrices--
have been used with qualitative success in the specification of nuclear 
densities in stable nuclei and the description of NA scattering for
projectile energies ($T_{Lab}$) below 1\,GeV. 
For $T_{Lab}$ above  this limit  we expect significant
dynamical changes due to dibaryonic fusion with subsequent fission
in the short range region of NN subsystems \cite{Fun01}.

Hitherto quite independently, several groups  successfully described
 intermediate energy NA scattering using  two nucleon 
$t$- or $g$-matrices as driving effective interactions \cite{Ray92}. 
Among them we distinguish two main philosophies under which a description
of NA collisions is made and which we specify as the
nonrelativistic Schr\"odinger approach  and as the relativistic Dirac 
approach \cite{Ray92,Amo00}. 
Common to both is the explicit use, and accurate treatment, of 
an interacting NN pair in the realm of other nucleons in a nucleus, and 
the need for an effective NN scattering amplitude  known 
and defined on- and off-shell.
For the more recent calculations of nonrelativistic NA optical 
potentials in momentum space 
\cite{Are90,Els90,Cre90}, a consistent treatment of the 
fully off-shell NN $t$- or $g$-matrices have been used.

As nearly all available NN 
potentials \cite{Lac80,Sto93,Sto94,Wir95,Mac01}
have been fitted to NN scattering data up to 350\,MeV, 
a nonrelativistic treatment of the full-folding optical model 
suffices most NA scattering applications consistent with the input NN model.
Recently, however, 
inversion potentials developed to fit elastic NN phase-shifts
for $T_{Lab}< 1.2$\,GeV have been used within the full-folding approach
for projectile energies as high as 500 MeV \cite{Are96}. 
Despite the fact that these were nonrelativistic applications, 
they allowed a better description of data relative to what was
found previously using 
traditional NN potential models. 
The lack of relativistic kinematics as well as pion production
and $\Delta$ excitations  within the NN pair did not manifest 
itself as a dramatic limitation of the model, 
even though the theoretical confidence level was reached. 
It is the primary aim of this work to remove these shortcomings and 
formulate a  momentum space full-folding model using {\em minimum-relativity}. 
With the appropriate modifications 
to the existing analysis programs we expect
to extend the confidence level into the GeV region and 
to obtain high quality 
full-folding optical model results at energies as high as 1.5\,GeV.

Extensive studies of NA scattering in the context of relativistic
Dirac models have been made \cite{Ray92,Mur87,Ott88,Tjo91,Coo93}.
While these models are closer to a fundamental formulation with the inclusion 
of relativistic kinematics and dynamics within a single 
framework, to date they all have used  fitted NN interactions that
do not  describe the phase shift data well.
This is a serious drawback as it is well established 
that any folding model requires a high quality NN interaction 
as input.

The description of NA scattering requires only modest nuclear matter 
densities and thus the need of NN $g$-matrices is less demanding. 
Furthermore only for projectile energies  below $500$\,MeV are 
nuclear medium effects  predominant with specific treatment of Pauli 
blocking and self consistent mean field effects being crucial. 
For medium and higher NN energies,  meson production and intrinsic 
hadron excitations  within the interacting pair are important.
Consequently  
the $t$-matrix no longer  is unitary in the elastic channel.
However as  
the low energy $t$- and $g$-matrices are well defined within potential
scattering theory,  we seek a continuation 
of NN potential models with an NN optical model potential (NNOMP).
We have devised and generated NNOMP 
for $0.3<T_{Lab}<3$ GeV and applied them to NA elastic scattering.  
The calculations are based upon a relativistically corrected full-folding 
optical model in momentum space that is an extension of a nonrelativistic 
predecessor \cite{Are90}.
Since the Lorentz contraction  scales as the ratio of the projectile
energy to its mass, it requires  nucleon projectile energies
above $400$\,MeV to have sizeable  contraction effects.
It is safe to include relativistic  kinematics in folding calculations 
when the projectile energy surpasses 300\,MeV. 
Relativistic kinematics  is widely used in pion-nucleus scattering
\cite{Tho80,Wei82,Aar68,Ern80,Gie82}.
An excellent  discussion on    some aspects about relativistic 
kinematics in NA scattering                        is given
in the review by Ray, Hoffmann and Coker \cite{Ray92}.

One of the important advantages of finding a nonrelativistic NA optical 
model potential is its well defined structure in terms of interacting 
nucleon pairs.
This framework has been remarkably successful in the study of low energy
scattering with its link to the nuclear shell model of single particle bound
states. 
The combination of target correlations and high quality NN
interactions  has provided a good first order description 
of the NA dynamics.
This nonrelativistic theoretical approach has been applied with
success for many energies and
targets and now it is timely to incorporate relativistic
corrections so that calculations can be made with energies 
as high as 1.5\,GeV.

A brief summary of the current full-folding optical model is given
in Section II and in which  the points where relativistic 
kinematic corrections are to be made are stressed. 
The current experimental situation of NN phase shift analysis up to 
3\,GeV is discussed in Section III.  
We describe also how the NNOMP are determined by the NN phase shift data.
In Section IV we set the framework of the $g$-matrix calculations 
and specify minimal relativity considerations.
In Section V some details of the full-folding calculations and
various applications are presented.
We discuss the role of the Fermi motion in the NN effective interaction
and analyze the sensitivity of NA scattering observables upon the use of
alternative approaches for the relativistic kinematics. 
We also examine total cross sections for nucleon elastic scattering 
and differential observables at beam energies from
the hundreds of MeV up into the GeV regime.
Section VI contains a summary and the main conclusions of our study.
Finally, we have included two Appendices.
In Appendix A the relativistic kinematics transformations  
in the context of the full-folding approach are
outlined as extracted from the articles by Aaron, Amado and 
Young \cite{Aar68} and by Giebink \cite{Gie82}.
In Appendix B we present the algorithm we have used to determine 
the NNOMP from data.


 \section{Full-folding framework}
In the nonrelativistic theory of the optical model potential,
the coupling between the projectile and the target nucleus in 
the elastic channel is given by the convolution between a two-body
effective interaction and the target ground-state mixed density.
In the projectile-nucleus c.m. frame, the collision of a projectile 
of kinetic energy $E$ is described by the optical potential $U(E)$ 
which in a momentum space representation is expressed as \cite{Are90}

\begin{equation}
\label{Uff}
U({\bf k}^\prime, {\bf k}; E) = \sum_{\alpha\leq\epsilon_F} 
\int \int d{\bf p}^\prime\;d{\bf p}\; 
\phi_\alpha^\dagger ( {\bf p}^\prime ) 
\langle {\bf k}^\prime {\bf p}^\prime 
| {\mathcal T}(\Omega_\alpha) | 
{\bf k} \; {\bf p} \rangle_{A+1} 
\phi_\alpha( {\bf p})
\end{equation}
where $\phi_\alpha $ represents target ground state
single-particle wave functions of energy $\epsilon_\alpha$,
and $\alpha\leq\epsilon_F$ restricts the sum to all levels 
up to the Fermi surface $\epsilon_F$.
The two-body ${\mathcal T}$ matrix is evaluated at
starting energies $\Omega_\alpha=m_p + E + m_t + \epsilon_\alpha$ 
consisting of the mass of the projectile $m_p$, its kinetic energy $E$,
the mass of the target nucleon $m_t$ and its 
binding energy $\epsilon_\alpha$.
The subscript A+1 indicates matrix elements in the projectile-nucleus c.m. 
frame and recoil effects have been neglected for simplicity.
In the most recent full-folding calculations of the optical potential
for NA scattering,
where explicit medium effects are incorporated in the two-body effective    
interaction, the ${\mathcal T}$-matrix is represented by 
an infinite nuclear matter $g$-matrix \cite{Are90}. 
In the absence of these medium modifications, the ${\mathcal T}$-matrix
has usually been approximated by the two-body scattering matrix 
associated with the collisions of two free particles \cite{Are90,Els90}.
None of these full-folding approaches incorporate the necessary 
relativistic kinematics needed for high energy processes.   

The above expression for the optical potential requires the 
${\mathcal T}$ matrix in the projectile-nucleus c.m. frame. 
However, most practical two-body potential models are designed 
to account for the scattering data in the two-body c.m., 
where a one-body wave equation (a Schr\"odinger kind equation)
is used to construct the realistic two-body bare potential
fit to elastic scattering and ground state data.
The practical problem that emerges then is how to make use of this
description to extract the needed effective interaction in the
projectile-nucleus c.m. with an adequate account of relativistic effects.
This has been a long standing problem in nuclear research and various
approaches have been proposed and discussed 
elsewhere \cite{Tho80,Wei82,Aar68,Ern80,Gie82}.
In the procedure followed here we retain the dynamical structure 
of the optical potential as expressed
by Eq. (\ref{Uff}) and identify, {\em via} a Lorentz boost, 
the corresponding kinematical variables involved in the 
two-body collision.
Thus, the transformation of the ${\mathcal T}$ matrix from the 
two-body (2B) to the projectile-nucleus (A+1) c.m. frame can be
done by considering  three separate aspects.
First, as Lorentz invariance of the flux is required,
an overall normalization factor --usually referred as the M\o ller 
flux factor-- mediates between the scattering amplitudes in the two frames. 
Second, the kinematics in the projectile-nucleus c.m. frame
needs to be transformed to the two-body c.m. system, which is the reference
frame where the bare potential model is defined.
And third, the transformation of the scattering matrix from the 
two-body to the projectile-nucleus c.m. frames involves the rotation
of the spins, an effect referred as the Wigner rotation.
This contribution has been studied in the context of the relativistic
`no-pair' potential for nucleon-nucleus scattering by 
Tjon and Wallace \cite{Tjo91} and was observed to yield 
rather moderate effects at nucleon energies between 200 and 500 MeV.
Although there is no statement about the importance of 
Wigner rotation contribution at the higher energies considered here, 
we shall neglect them in the present work.

\begin{figure}
\scalebox{0.6}{\includegraphics{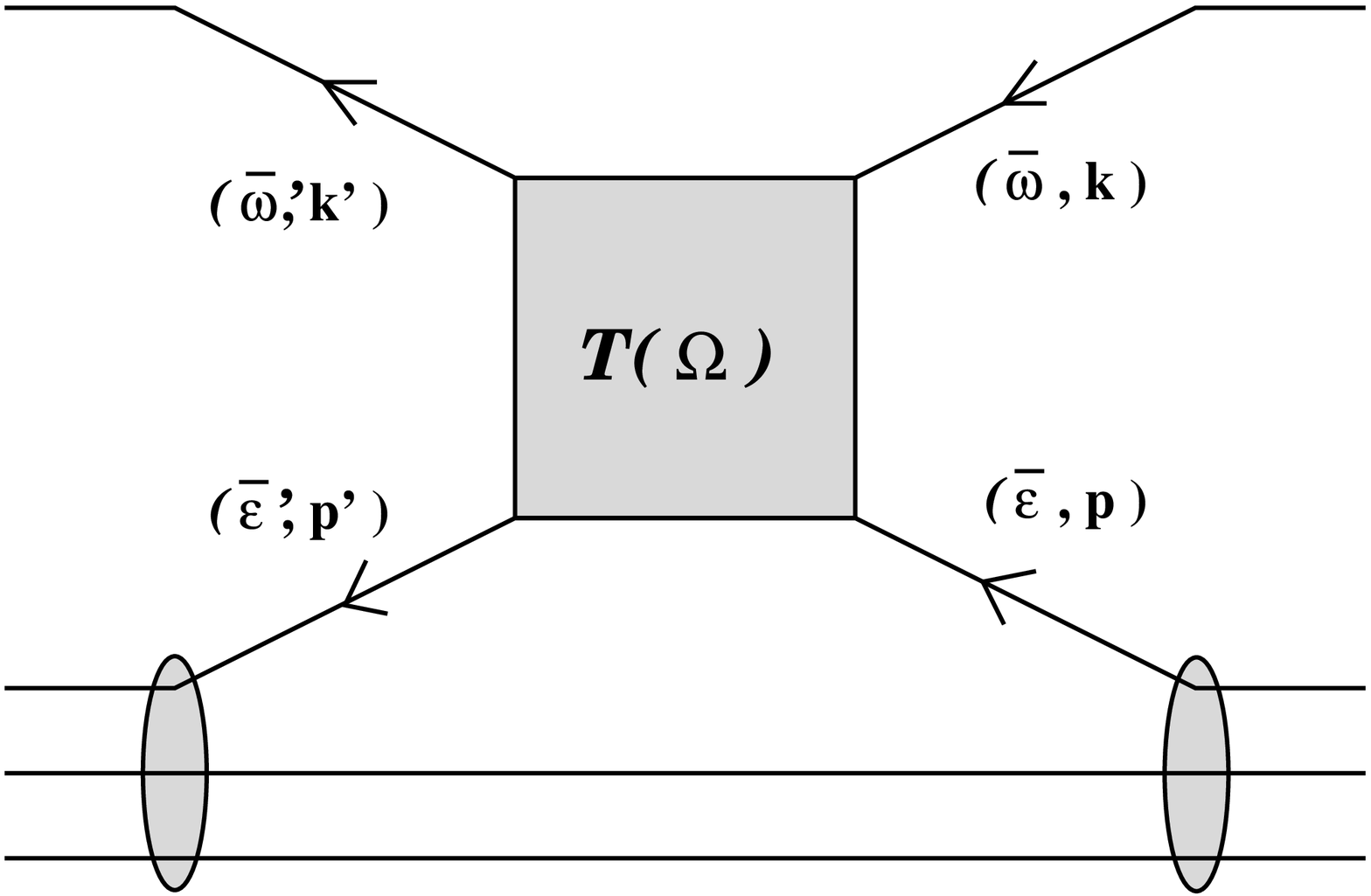}}
\caption{\label{tmatrix}
         Schematic representation of the collision of the projectile with
         a target nucleon.
         Quantities in parentheses represent the four-momenta
         of the colliding nucleons.
        }
\end{figure}

Consistent with the notation introduced in Fig.~\ref{tmatrix} for the 
coupling between the projectile and the target struck nucleon,
we denote their respective incoming and outgoing four-momenta
\begin{eqnarray}
\label{4mtum}
k &=& (\bar\omega,{\bf k} ), \qquad 
k^\prime = (\bar\omega^\prime,{\bf k}^\prime) \;\; ; \nonumber \\
p &=& (\bar\varepsilon,{\bf p} \; ), \qquad
p^\prime =(\bar\varepsilon^\prime,{\bf p}^\prime) \;\; .
\end{eqnarray}
When translational invariance is assumed, such is the case of free   
or particles in infinite nuclear matter, the two-body interaction is characterized by the
starting energy $\Omega$
and the total momentum of the colliding particles
${\bf Q}$,
\begin{equation}
\label{xform}
\left \langle {\bf k}^\prime {\bf p}^\prime 
| {\mathcal T}(\Omega ) | 
{\bf k} \; {\bf p} \right \rangle_{A+1} = 
\eta({\bf k}^\prime {\bf p}^\prime ;{\bf k} \; {\bf p} ) \;
\left \langle {\bf k_r}^\prime,-{\bf k_r}^\prime
| \tau_{\bf Q }(\sqrt{s} )|
{\bf k_r} , -{\bf k_r}\right \rangle_{2B} \;
\delta ( {\bf Q}^\prime - {\bf Q} ) \; .
\end{equation}
Here the Dirac delta-function makes explicit the total three-momentum 
conservation of the two-body collision, ${\bf Q}^\prime ={\bf Q}$, where
\begin{equation}
\label{Qtot}
{\bf Q}  = {\bf k} + {\bf p} \;\; ,\quad
{\bf Q}' = {\bf k}^\prime + {\bf p}^\prime , \hfill
\end{equation}
and $\sqrt{s}$ represents the energy in two-body c.m., 
\begin{equation}
\label{s_com}
s = \Omega^2 - {\bf Q}^2\; .
\end{equation}
The overall coefficient $\eta$ is the M\o ller flux factor,
\begin{equation}
\label{moller}
\eta({\bf k}^\prime {\bf p}^\prime ;{\bf k} \; {\bf p} ) = 
\left [
\frac{\omega({\bf k_r}^\prime) \varepsilon(-{\bf k_r}^\prime) 
      \omega({\bf k_r})        \varepsilon(-{\bf k_r}) }
{     \omega({\bf k}^\prime)  \;     \varepsilon({\bf p}^\prime)\; 
      \omega({\bf k})      \;        \varepsilon({\bf p})}
\right ]^{1/2} \;\; ,
\end{equation}
where the energies $\omega$ and $\varepsilon$ 
are on-mass-shell, i.e. $\omega^2=m_p^2 + {\bf k_r}^2$ and
$\varepsilon^2=m_t^2 + {\bf k_r}^2$.

What remains to be specified is a Lorentz transformation for the relative 
momenta ${\bf k_r}$ and ${\bf k_r}^\prime$. 
To this purpose we followed the approaches 
introduced by Aaron, Amado and Young (AAY) \cite{Aar68}
and Giebink \cite{Gie82}.
Although, both assume a representation of the scattering matrix in the 
two-body c.m., they differ in the way the Lorentz boost is devised.
Details are given in Appendix A  where we show that in both cases  
the relative momenta can be cast into the form 
\begin{eqnarray}
\label{rel_mtum}
{\bf k_r} &=& W {\bf k} - (1-W){\bf p}\; ,  \nonumber\\
{\bf k_r}^\prime &=& W^\prime {\bf k}^\prime-(1-W^\prime){\bf p}^\prime\; , 
\end{eqnarray}
with $W$ and $W^\prime$ being functions of the momenta of the colliding 
particles (Eqs. (\ref{k_gie}) or (\ref{k_aay})). 
At low beam energies both prescriptions meet the nonrelativistic
limit $W\sim W'\sim m_p/(m_p+m_t)$, as expected.


 \section{The NN optical model potential}
Of the whole spectrum, low and medium energy NN scattering traditionally
is  described in terms of Hermitian potentials.
At medium energies, production processes and inelasticities
become possible and several elementary systems composed of nucleons
and mesons contribute to NN scattering. 
At present there is no high quality description of NN  scattering above 
the inelastic threshold either in terms of QCD or
in terms  of  nucleons and mesons. 

A high quality fit of on-shell $t$-matrices by means of a potential model 
is very desirable as it provides  extensions of the effective 
interaction into the off-shell domain and into a nuclear medium, 
which are important dynamical features in few and many 
body calculations.
Many examples using microscopic optical
model potentials for elastic nucleon-nucleus scattering and
bremsstrahlung reactions, have shown that it is crucial to 
have on-shell $t$-matrices in the best possible agreement with NN  
data at all energies. 
Concomitantly one needs high precision NN  data against which one can 
specify NN interactions.
To this purpose we have relied on a large body of experimental NN data 
whose parameterization in terms of amplitudes and phase shifts are 
smooth for energies to 3 GeV \cite{GWU00}. 
This is a supposition for the construction of an NN potential 
above 300 MeV.

There are many studies of few and many body problems
in the low energy regime  $T_{Lab} < 300$\,MeV and the results
have consequences for any model extension above threshold.
We note in this context that significant off-shell
differences in  $t$-matrices are known to exist among   
the theoretically well motivated boson exchange models of NN
scattering. It remains difficult to attribute
with certainty any particular dynamical or kinematical feature with those
differences. Non-locality, explicit energy dependence and features  associated
with relativistic kinematics are some possibilities.
In contrast, there is the quantum inverse scattering approach by
which  on-shell $t$-matrices can be continued into the off-shell domain.
A specific method is the Gel'fand-Levitan-Marchenko
inversion algorithm for Sturm-Liouville equations.
This approach to specify $t$-matrices off-shell is appropriate when the
physical S-matrix is unitary and the equation of motion is of the
Sturm-Liouville type. Such  is valid without modification  for
NN  $t$-matrices in the energy regime below 300 MeV, and for the
unitary part of the S-matrix above that energy.

In the spirit of general inverse problems,
we have extended the available low energy  potential by additional
complex potential terms
which are determined from a perfect reproduction of
the experimental data  (here the partial wave phase
shift analysis) for all energies above 300 MeV. By that means
NN  optical models were generated,  separately for each partial wave.
The  algorithm we  have developed allows studies of
complex local and/or separable potentials in combination with
any  background  reference potential \cite{Ger98,Fun01}. 
Here we limit the reference
potential  to the well known real coordinate space
potentials from Paris \cite{Lac80}, Nijmegen \cite{Sto94}
(Reid93, Nijmegen-I, Nijmegen-II), Argonne \cite{Wir95} (AV18),
and from inversion \cite{San97}.
To them we add channel-dependent complex separable potentials with
energy dependent strengths \cite{Fun01}. 
For a given input data set there is a unique
NNOMP within a given potential class.
Some of these issues are outlined in Appendix B.

NN scattering is a long standing problem
which  has been reviewed often as the database developed.
The low energy data has been analyzed by the VPI/GWU group
\cite{GWU00} for $T_{Lab}\leq 400$ MeV, the  Nijmegen group 
with the  NN phase shift results  PWA93 for
$T_{Lab}\leq 350$ MeV, and by Machleidt \cite{Mac01} giving the
Bonn-CD-2000.
Of these, the VPI/GWU group has given many solutions over the years, 
the latest are for energies  to 3 GeV \cite{SAID}.
 We have  used the solutions SP00, FA00 and
WI00 in our calculations and found results that differ but marginally.
Thus hereafter in the main we refer solely to the results of calculations
based upon SP00.

As with the NN phase shift analysis, one boson exchange potentials have 
received several
critical reviews \cite{Mac01}, including observations
that there are small variations between phase shift analysis
and potential models below the subthreshold
domain $T_{Lab} <300$ MeV. A theoretically stable result
would require many quantities, that need be specified {\em a priori},
to be  determined by independent sources. At present that does
not seem feasible and all current potentials rely upon
fits of many of their parameters to the same data. All such fits,
however, have been made independently of each other and are
based upon differing theoretical specifications of
the boson exchange model dynamics.

Above 300 MeV, reaction channels open and
the elastic channel S-matrix no longer is unitary.
Only the $\Delta$(1232) resonance has a
low energy threshold and a relative small width of 120 MeV.
Therefore it is the
only resonance we expect to be obviously visible in the energy variation of
the elastic scattering phase shifts. In particular one
notices typical variations
in the $^1D_2$, $^3F_3$, and $^3PF_2$ channels. Otherwise the phase shifts
to 3 GeV vary smoothly as functions of  energy.
Together with the strong spin-isospin coupling, this property infers
optical potentials that are  channel dependent in
contrast to the NA case for which assumed central and spin-orbit
potentials are partial wave independent.
The plethora of reaction channels that open  to 3 GeV, and the requirement
of an NN optical potential prescription, 
mean that it is
an interesting  task for a microscopic model
to  link QCD substructures to NN scattering
phase shift functions in analogy to that successful prescription by
which NA optical potentials have
been determined by folding effective interactions.

To describe this developing system for $0.3<T_{Lab}<3$ GeV
we  used Feshbach theory to specify the
optical potential. An important feature of
that theory is the projection operator formalism with $P$  and
$Q$  subspaces, which divide the complete
Hilbert space $(P+Q)$ into the elastic scattering channel,
the $P$ space, and the inelastic and reaction channels, the $Q$ space. 
This infers a complex and separable
component in the optical potential with an energy dependent strength.
If a very large number of intermediate states contribute, the effect equates to
a local potential operator. 

A covariant description of NN scattering formally is given
by the Bethe-Salpeter equation
\begin{equation} \label{eqn_III.3}
{\cal M} = {\cal V} + {\cal V}{\cal G}{\cal M}\ ,
\end{equation}
where  $\cal M$ are invariant amplitudes
that are based upon all connected two particle irreducible diagrams.
This equation serves generally as an ansatz for
approximations. Of those, the three dimensional  Blankenbecler-Sugar reduction
is popular and sufficient for our purpose to  define an NN potential
\cite{Mac01}.
The amplitudes are now expressed with the reduced terms
and they  satisfy a three-dimensional equation
\begin{equation}\label{eqn_III.8}
{\cal M} ({\bf q}^\prime,{\bf q}) = {\cal V} ({\bf q}^\prime,{\bf q}) +
\int { \frac{d^3k}{(2 \pi)^3} }{\cal V} ({\bf q}^\prime,{\bf k})
 { \frac{M^2}{E_{\bf k}} }
\frac{\Lambda^+_{(1)} ({\bf k}) \Lambda^+_{(2)}(-{\bf k})}{ {\bf q}^2 - {\bf k}^2 + i \varepsilon}
{\cal M} ({\bf k},{\bf q}).
\end{equation}
Taking matrix elements with only positive energy spinors,
an equation with  minimum relativity results
for the  NN $t$-matrix, namely
\begin{equation}\label{eqn_III.9}
{\cal T} ({\bf q}^\prime,{\bf q}) =  {\cal V} ({\bf q}^\prime,{\bf q}) +
\int { \frac{d^3k}{(2 \pi)^3}}  {\cal V} ({\bf q}^\prime,{\bf k})
 { \frac{M^2}{E_{\bf k}}} \frac{1}{{\bf q}^2 - {\bf k}^2 + i \varepsilon}
{\cal T} ({\bf k},{\bf q}).
\end{equation}
Using  the substitutions
\begin{equation}\label{eqn_III.10}
T ({\bf q}^\prime,{\bf q}) = \left( \frac{M}{E_{\bf q^\prime}}
\right)^{\frac12}
{\cal T} ({\bf q}^\prime,{\bf q})
\left( \frac{M}{E_{\bf q}} \right)^{\frac12}
\end{equation}
and
\begin{equation}\label{eqn_III.11}
V ({\bf q}^\prime,{\bf q})
= \left( \frac{M}{E_{\bf q^\prime}} \right)^{\frac12}
{\cal V} ({\bf q}^\prime,{\bf q})
\left( \frac{M}{E_{\bf q}} \right)^{\frac12},
\end{equation}
a simplified form of the $t$-matrix is obtained. It is the familiar
Lippmann-Schwinger equation
\begin{equation}\label{eqn_III.12}
 T ({\bf q}^\prime,{\bf q}) =  V ({\bf q}^\prime,{\bf q}) +
\int { \frac{d^3k}{(2 \pi)^3} }  V ({\bf q}^\prime,{\bf k})
 {\frac{M}{{\bf q}^2 - {\bf k}^2 + i \varepsilon}}
 T ({\bf k},{\bf q})\ .
\end{equation}
The strategic importance of this result is that it defines a sensible
continuation of a $t$-matrix, constrained on-shell by the experimental 
data and phase shift analysis, into the off-shell domain as required 
by the full-folding optical model.
Thus we do not rely primarily on a fundamental theoretical 
result but rather on 
experimental NN data and the moderate sensitivity of NA scattering 
to alternative off-shell continuations. 
It is an important result also of this analysis
that on-shell equivalent NN optical model potentials  yield very similar
NA scattering  observables -- irrespectively of differences in the
off-shell domain and the constraint off-shell
continuation defined with Eqs.\,(\ref{eqn_III.8}-\ref{eqn_III.12}).


 \section{In--medium effective interaction}
A crucial step in the description of NA scattering processes has
been  the definition of an effective interaction based on the bare 
two-body interaction in free space.
Such has been the philosophy of the early (local) folding models and 
the most recent nonrelativistic full-folding models \cite{Are90}.
The effective interaction, $\cal T$ in Eq. (\ref{Uff}), is obtained  
in the framework of 
Brueckner-Bethe-Goldstone for the $g$-matrix.
The extension of this approach to high energy applications requires a minimal
account of relativistic corrections.
Along this line we have followed the discussion by Brockmann and 
Machleidt \cite{Bro90}, where a relativistic three-dimensional reduction  
of the Bethe-Salpeter equation is used to describe the 
interaction between nucleons in the nuclear medium. 
If only matrix elements between positive-energy spinors are taken,
then the medium-modified invariant amplitude in an arbitrary frame
reads (cf. Eqs. (A17, A18) in Ref. \cite{Bro90})
\begin{equation}
\label{thompson}
{\cal G}_{\bf Q}({\bf q}',{\bf q}; s) = {\cal V}_{\bf Q}({\bf q}',{\bf q})
+     \int \frac{{d^3k}}{(2\pi)^3} {\cal V}_{\bf Q}({\bf q}',{\bf k})
\left (\frac{M}{E_{\frac{1}{2}{\bf Q}+{\bf k}}} \right )
\frac{M\bar Q({\bf Q};{\bf k})}
     {\frac{1}{4}s + \frac{1}{4}{\bf Q}^2 - 
E_{\frac{1}{2}{\bf Q}+{\bf k}}^2 + i\varepsilon}
{\cal G}_{\bf Q}({\bf k},{\bf q}; s)\;.
\end{equation}
Here the momentum ${\bf Q}$ represents the momentum of the pair with respect
to the background, and $\bar Q$ the Pauli blocking operator           
which projects onto unoccupied intermediate states.
For the above expression angle averages have been used, i.e.
$|\frac{1}{2}{\bf Q}+{\bf k}|^2\approx \frac{1}{4}{\bf Q}^2 + {\bf k}^2$,
and the $s$ invariant has been defined as 
$s=4E_{\frac{1}{2}{\bf Q}+{\bf q}}^2-{\bf Q}^2$.
This approach, in the context of the Dirac-Brueckner-Hartree-Fock 
approximation, has been applied with reasonable success to the study 
of infinite nuclear matter \cite{Bro90} as well as finite nuclei ground 
state properties \cite{Mut90}.

An appealing feature of the above equation for ${\cal G}$ is its direct 
connection with the bare NN potential model in free space. Indeed, 
adopting the same definitions as in Eqs.~(\ref{eqn_III.10}) and (\ref{eqn_III.11}), 
\begin{equation}
\label{def_t}
g_{\bf Q}({\bf q}',{\bf q}; \sqrt{s})=        \sqrt{\frac{M}{E_{\bf q'}}}
  {\cal G}_{\bf Q}({\bf q}',{\bf q}; s)\sqrt{\frac{M}{E_{\bf q }}} \;,
\end{equation}
and
\begin{equation}
\label{def_v}
V({\bf q}',{\bf q})=        \sqrt{\frac{M}{E_{\bf q'}}}
  {\cal V}({\bf q}',{\bf q})\sqrt{\frac{M}{E_{\bf q }}} \; ,
\end{equation}
the following equation for the $g$-matrix is obtained
\begin{equation}
\label{gMat}
g_{\bf Q}({\bf q}',{\bf q}; \sqrt{s}) =  V_{\bf Q}({\bf q}',{\bf q})
+     \int \frac{d^3k}{(2\pi)^3}  V_{\bf Q}({\bf q}',{\bf k})
\left (\frac{E_{\bf k}}{E_{\frac{1}{2}{\bf Q}+{\bf k}}}\right )
  \frac{M\bar Q({\bf Q,{\bf k}})}
{\frac{1}{4}s + \frac{1}{4}{\bf Q}^2 - 
E_{\frac{1}{2}{\bf Q}+{\bf k}}^2 + i\varepsilon}
g_{\bf Q}({\bf k},{\bf q}; \sqrt{s})\; .
\end{equation}

The above Dirac-Brueckner approach differs in a non-trivial way from the
conventional nonrelativistic Brueckner approach. 
The density dependence of the one-boson-exchange interaction by means 
of effective Dirac spinors and the explicit relativistic
kinematics are features which have no counterpart in the traditional
Brueckner approach. 
However, with the suppression of these relativistic dynamical effects 
we can extract a minimum of relativistic features needed in the 
nonrelativistic model.
In our approach selfconsistency is demanded with the following choice 
for the quasi-particle spectrum,
\begin{equation}
\label{spectrum}
E^2_{\bf p}  = {\bf p}^2 + \left(M+U({\bf p})\right)^2\; .
\end{equation}
As in the usual Brueckner-Bethe-Goldstone approach, the quasi-potential
$U({\bf p})$ is obtained selfconsistently with the use of the continuous  
choice at the Fermi surface. 
As a first check, we consider  the case of two nucleons 
interacting in free space, for which the 
Pauli blocking operator becomes the identity and 
the nuclear selfconsistent field vanishes. 
Furthermore, if the interaction is described in the pair c.m. (${\bf Q}=0$) 
then the $g$-matrix corresponds to the free scattering matrix $T$ as 
described by the BbS equation (cf. Eq. \ref{eqn_III.12}). 
This limit is immediately verified upon substitution of $\sqrt{s}$ by 
$2\sqrt{q_\circ^2 + M^2}$,  with $q_\circ$ the on-shell c.m. relative momentum. 
Thus,
\begin{equation}
\label{BbS}
T({\bf q}',{\bf q}; \sqrt{s}) =  V({\bf q}',{\bf q})
+     \int \frac{d^3k}{(2\pi)^3}  V({\bf q}',{\bf k})
  \frac{M}{q_\circ^2 - k^2 + i\varepsilon} T({\bf k},{\bf q}; \sqrt{s})\; ,
\end{equation}
which is the nonrelativistic Lippmann-Schwinger equation with 
the pole at the relativistically correct momentum.

The other case of interest is infinite nuclear matter, where
the relativistic structure of the quasi-particle 
spectrum introduced in Eq. (\ref{spectrum}) can be assessed considering 
the following power expansion in terms of $U/M$ 
\begin{equation}
\label{e_appr}
E^2_{\bf p}/M  \approx 2 \left ( \frac{{\bf p}^2}{2M} + U({\bf p})\right )
+ M ( 1 + {\cal O}[ (U/M)^2 ])\;.
\end{equation}
If we substitute ${\bf p}^2$ by the angle averaged quantity
$\frac{1}{4}{\bf Q}^2+{\bf k}^2$ and proceed similarly with $s$,
 the similarity of the energy denominator in 
Eq. (\ref{gMat}) with the one obtained using the nonrelativistic 
propagator is evident.
An estimate of the accuracy of the above approximation 
at normal densities can be made considering $(U/M)\lesssim$ 1/10.
In such a case the above form of $E^2$ yields a propagator equivalent 
to its nonrelativistic counterpart with an accuracy better than 1\%.
This result supports the use of the nonrelativistic selfconsistent
scheme to obtain the quasi-potentials.

With the above considerations, the calculation of the two-body effective  
interaction needed in the full-folding optical potential proceeds with
the proper choice of the $s$-invariant in Eq. (\ref{gMat}), and 
consistent with the starting energy $\Omega_\alpha$ 
defined in Eq. (\ref{Uff}). 
Since $\Omega_\alpha$ represents the total pair energy of interacting
nucleons with total momentum ${\bf Q}$, the $s$-invariant is 
simply given by $s=\Omega_\alpha^2-{\bf Q}^2$.


 \section{Applications}
A few remarks about the actual implementation
of the computational procedures, aimed to obtain the full-folding
optical potential $U(E)$, are worthy to note.
First of all, we use the Slater approximation for the nuclear mixed density
$\phi^\dagger\phi$ in Eq. (\ref{Uff}). 
This approximation has been discussed in the past \cite{Are90} and becomes 
particularly suitable for the present applications, 
where we rely on the same point nuclear densities for the target ground state. 
Furthermore, we make use of the infinite nuclear matter $g$-matrix to 
represent the effective interaction between the projectile and the
target nucleon \cite{Are90}. 
As a result, the $g$-matrix evaluated at a Fermi 
momentum $k_F$ is folded with the target ground state density $\rho$ 
at a local momentum $\hat k(R)$.
The Slater approximation suggests the  ansatz $k_F=\hat k$, where
the local momentum $\hat k$ is determined from the density by
$\hat k^3(R)=3\pi^2\rho(R)/2$.
These considerations yield for the full-folding optical model 
potential
\begin{equation}
\label{Uff_rho}
U({\bf k}', {\bf k}; E) = 4\pi\int d{\bf R} \;
e^{i({\bf k}' -{\bf k})\cdot {\bf R}} \left (
\rho_p({\bf R})\; \bar g_{pN}({\bf k}',{\bf k})  +  
\rho_n({\bf R})\; \bar g_{nN}({\bf k}',{\bf k})
\right ) \; ,
\end{equation}
where $\rho_p$ and $\rho_n$ are the local proton and neutron point
densities respectively, and $\bar g_{NN}$ represent the  off-shell
Fermi-averaged amplitudes in the NN channel.
For a particular channel this amplitude depends on the 
nuclear matter density {\em via} $k_F$ (implicit in $g$), 
and the local momentum $\hat k$ which sets the bounds for the Fermi 
motion of the target nucleons. 
This is expressed as
\begin{equation}
\label{g_on}
\bar g_{NN}({\bf k}',{\bf k}) =
\frac{3}{4\pi \hat k^3} \int
\Theta(\hat k - |{\bf P}| )
g_{{\bf K}+ {\bf P}} ( {\bf k_r}', {\bf k_r}; \sqrt{s} ) \; d{\bf P}\; ,
\end{equation}
where ${\bf K}= ({\bf k}+{\bf k}')/2$ and
the relative momentum ${\bf k_r}$ and ${\bf k_r}^\prime$ are
obtained following each of the two relativistic prescriptions 
discussed in Appendix A. 
We stress at this point that these amplitudes are calculated fully off-shell 
and that no assumption is made regarding the coordinate space structure
of the $g$-matrix. 
With these considerations the full-folding optical potential
becomes a genuine non-local operator. 
Its use in the Schr\"odinger equation involves integro-differential 
equations which are solved exactly within numerical accuracy.

The nuclear matter calculations for the $g$-matrix were done with 
fully self-consistent fields at various values of $k_F$. 
We have considered the inversion potentials based upon the {\sc sp00} 
phase-shifts solution with  Nijmegen-I, -II and Reid-93 reference potentials
 ({\sc sp00-nij1}, {\sc sp00-nij2} and {\sc sp00-re93} respectively),
Argonne reference potential ({\sc sp00-av18}), Paris reference potential
({\sc sp00-pari}) and 
Gel'fand-Levitan-Marchenko inversion reference potential ({\sc sp00-invs}).

\subsection{Medium and Fermi motion effects}

\begin{figure}
\scalebox{0.6}{\includegraphics{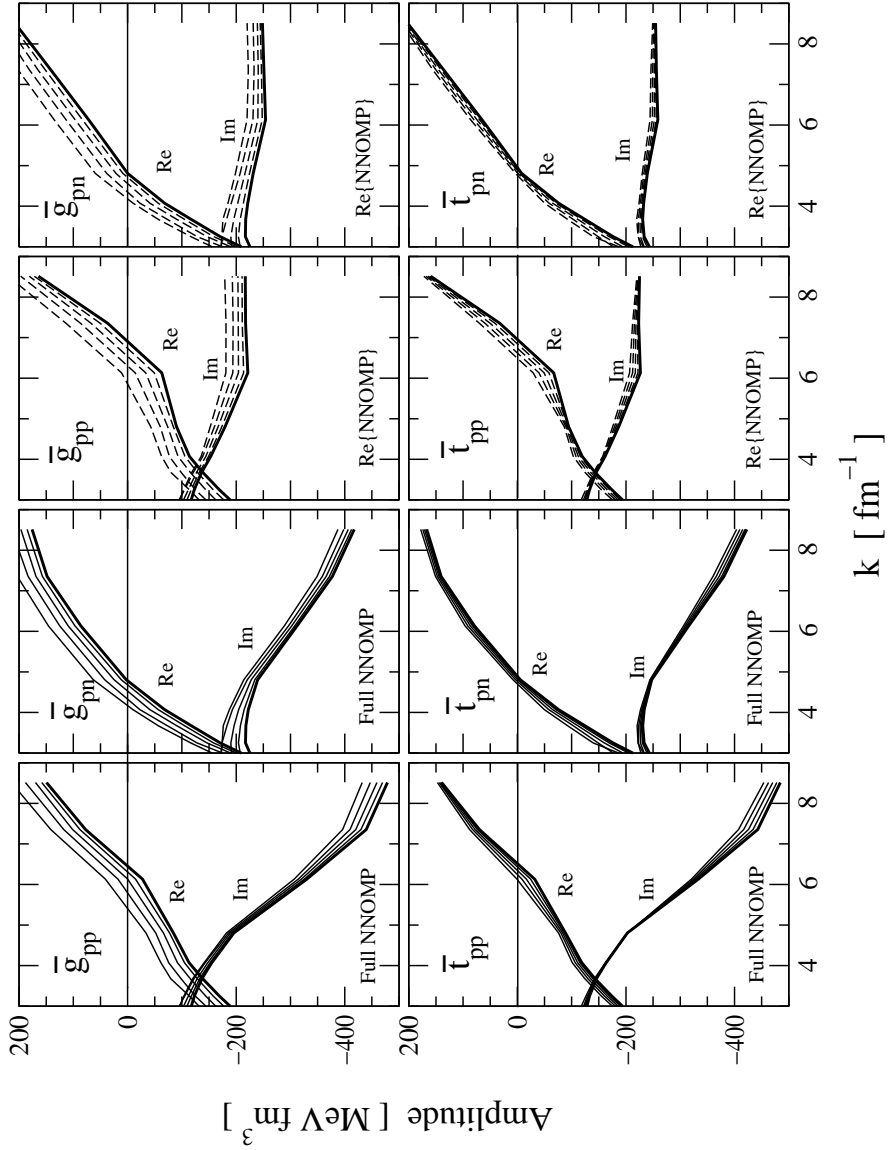}} 
\caption{\label{g_av18}
        The Fermi-averaged forward amplitude 
        $\bar g_{NN}$ at $k_F$ = 1 fm$^{-1}$ (upper frames) and
        $\bar t_{NN}$ (lower frames) based on the {\sc sp00-av18} 
        NNOMP at local momenta $\hat k$=0.6(0.2)1.4 fm$^{-1}$ in the 
        $pp$ and $np$ channels, and as functions of the projectile momentum. 
        The four leftmost frames represent results using the full NNOMP
        whereas the four rightmost frames with the dashed curves correspond 
        to results whith the imaginary part of the NNOMP suppressed.
        The thick solid curves denote results for $\hat k$=0.6 fm$^{-1}$.
        }
\end{figure}

To disclose some of the features exhibited by the Fermi-averaged 
effective interaction in the context of the full-folding approach, we 
analyze the $\bar g$ averages in the proton-proton ($pp$) and 
neutron-proton ($np$) channels. 
Here we focus on the on-shell
forward matrix element. 
In this case the $\bar g_{NN}$ amplitude depends on the NA projectile
momentum $|{\bf k}|$, the Fermi momentum $k_F$, 
and the local momentum $\hat k$. 
In the case of the free $t$-matrix approach for the interaction
we set $k_F=0$ but allow for the variation of ${\bf P}$ implied by the  
local nuclear density to account for the Fermi motion in the nucleus 
($|{\bf P}|\leq \hat k$).
In the case of a $g$-matrix element we set $\hat k=k_F$.
In Fig. \ref{g_av18} we show the real and imaginary components of
$\bar g_{pp}$ and $\bar g_{np}$, based on the {\sc sp00-av18} NNOMP,
as functions of the beam momentum and for the sequence 
$\hat k=0.6(0.2)1.4$ fm$^{-1}$.
In each frame we draw the case $\hat k=0.6$ fm$^{-1}$ with a thick solid 
curve; the following values of $\hat k$ depart sequentially from this 
reference curve.
The upper and lower frames correspond to $g$- and $t$-matrix results 
respectively.
To assess the role of the imaginary part of the NNOMP,
the results based on the full model are 
shown in the four leftmost frames with solid curves while the
results where the imaginary part is suppressed are
shown in the four rightmost frames with dashed curves.
The two relativistic kinematics prescriptions --Giebink and AAY--
yield almost indistinguishable results for the amplitudes. 

A  comparison of results in  the upper and lower frames 
indicates more dispersion due to the Fermi motion within the $g$-matrix
than in the $t$-matrix approach. This is a 
feature which is more pronounced in the real than the absorptive 
component of the amplitude.
This is a clear indication of the role of the self-consistent fields 
in the Fermi-averaged quantities.
In the context of the full NNOMP
however, the manifestation of this sensitivity
becomes diminished at projectile energies 
above 500 MeV as the real component of all amplitudes for $|{\bf k}|$
above $\sim$5 fm$^{-1}$ are significantly smaller than their
imaginary counterparts.
Such is not the case at the lower energies, where both the real
and imaginary components of the amplitude become comparable. 
Thus, we do expect more sensitivity under Fermi motion in the 
context of the $g$ matrix at projectile energies below 500 MeV. 

Another feature that emerges from Fig. \ref{g_av18} is the
sensitivity of the absorptive part of $\bar g$ 
to the presence of the imaginary part of the NNOMP.
Indeed, when this part is suppressed, the absorptive component 
of the amplitude saturates above $\sim$5 fm$^{-1}$,
in contrast with the full NNOMP where the trend of this 
absorption is to increase.
The manifestation of this feature becomes clear when the Fermi
average enters fully off-shell in the evaluation of the full-folding
potential, as discussed in the following section.

\subsection{Total cross sections}

\begin{figure}
\scalebox{0.60}{\includegraphics{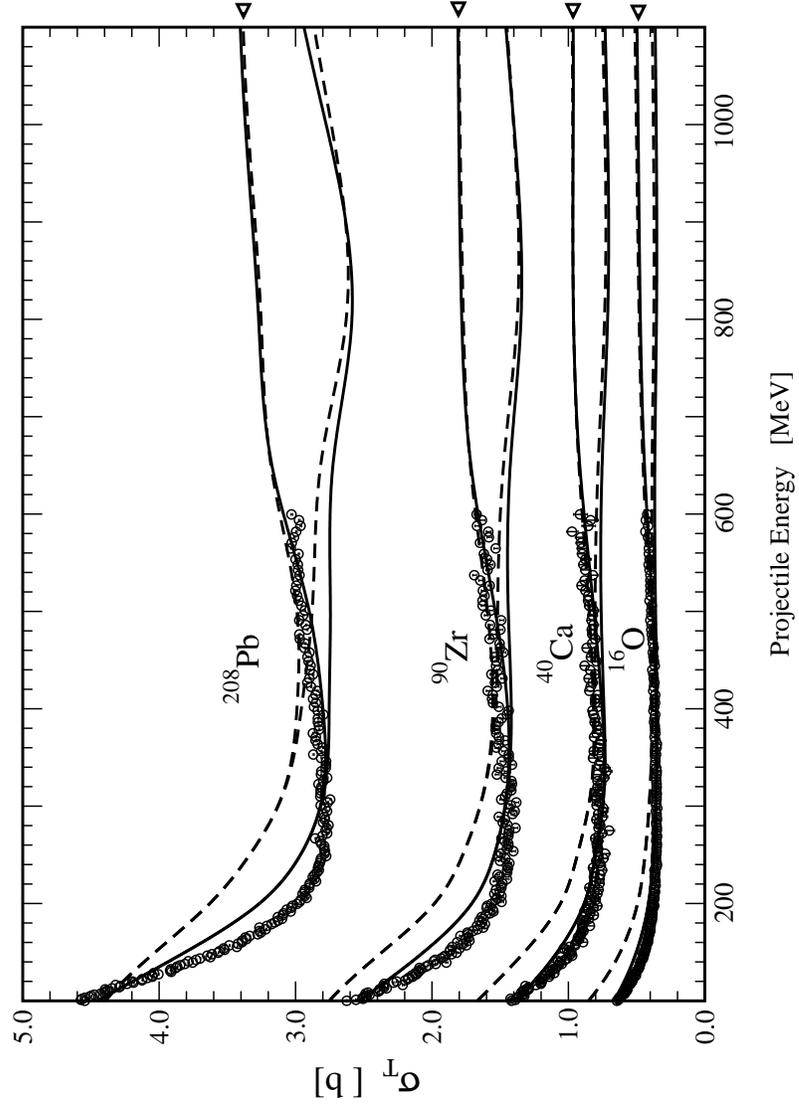}} 
\caption{\label{XTotalX4} 
Total cross section for neutron elastic scattering from
$^{208}$Pb, $^{90}$Zr, $^{40}$Ca and $^{16}$O as functions 
of the projectile energy.
The data \cite{Fin93} are represented with open circles.
The solid and dashed curves represent full-folding results using the
$g$- and $t$-matrix respectively. 
The curves corresponding to the full NNOMP are marked with a triangular 
label at their right end, 
whereas those results with the imaginary part of the NNOMP 
suppressed are unmarked.
}
\end{figure}

A global assessment of the full-folding model to the 
inclusion of relativistic kinematics and features of the underlying
NN potential model over a wide energy range is made first
by studying total cross sections for neutron-nucleus
elastic scattering.
In Fig. \ref{XTotalX4} 
we show the measured \cite{Fin93} and calculated total cross sections 
for neutron elastic scattering from $^{16}$O, $^{40}$Ca, $^{90}$Zr
and $^{208}$Pb at beam energies ranging from 100 MeV up to 1 GeV. 
These cross sections are obtained by solving the scattering equations
using full-folding optical potentials calculated as in Ref. \cite{Are90},
modified to include the kinematics discussed in this work. 
For completeness in this comparison we have considered the full NNOMP 
within the $g$- (solid curves) and $t$-matrix (dashed curves) approaches. 
In order to assess the role of the absorptive contribution of the NNOMP,
we have also included results suppressing the imaginary part.
These results are shown unmarked, whereas those using the full NNOMP
have been labeled with a triangle at their right end.
The data are represented with open circles.
From this figure we observe a remarkable agreement between
the full NNOMP $g$-matrix results and the data, particularly at energies
above $\sim$200 MeV. 
 Such is not the case for the $t$-matrix approach, or when 
the imaginary part of the bare NN potential
is suppressed. 
In the former case the lack of nuclear medium effects becomes 
 pronounced for   projectile energies as high as 500 MeV in the case 
of Pb, but less for lighter nuclei.
Conversely, the imaginary part of the NNOMP is crucial 
for the adequate description of  cross section data
at beam energies above 400 MeV, 
as observed when comparing the labeled and unlabeled curves.

\begin{figure}
\scalebox{0.6}{\includegraphics{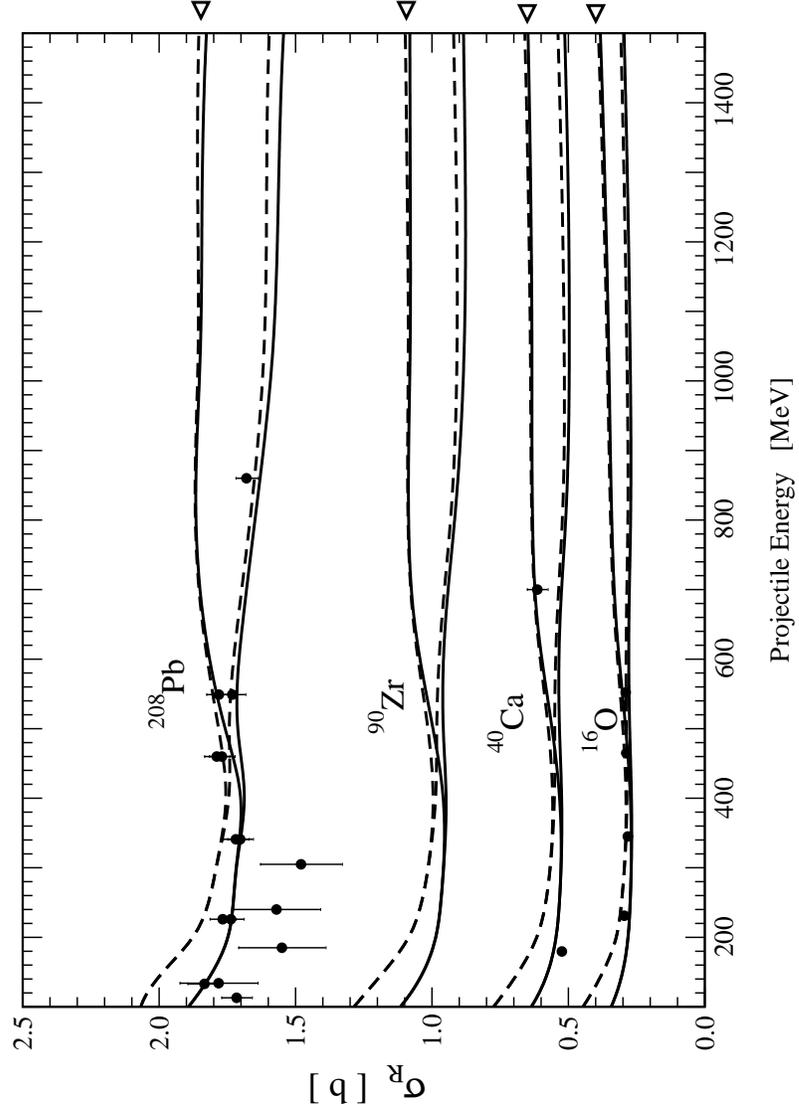}} 
\caption{\label{XReactX4}
Predicted reaction total cross section for proton elastic scattering from
$^{208}$Pb, $^{90}$Zr, $^{40}$Ca and $^{16}$O as functions of the beam energy.
The data were taken from Ref. \cite{Car96}.  
The solid and dashed curves represent full-folding results using the
$g$- and $t$-matrix respectively. 
The curves corresponding to results based on the full NNOMP have been
marked with a triangular label at their right end, 
whereas those results with the imaginary part of the NNOMP 
suppressed are unmarked.
}
\end{figure}

Total reaction cross sections for proton-nucleus  elastic scattering
at high energies are of increasing interest, 
particularly with current trends using spallation facilities 
with high energy beams. 
Thus, we calculated total reaction cross-section for elastic
proton scattering, at beam energies up to 1.5 GeV,
and compare them with available data \cite{Car96}.
In strict analogy with the previous applications, and use 
of notations as in Fig. \ref{XTotalX4},  in
Fig. \ref{XReactX4} we show  $g$- (solid curves) and $t$-matrix (dashed curves) 
results wich 
used the full  or  suppressed the imaginary part of the NNOMP.
Here again we have marked with triangular labels those results based on the 
full NNOMP,
and left  unmarked the ones      with the imaginary part suppressed.
The data  are  represented by open circles.
The results shown in Fig. \ref{XReactX4} used  
the {\sc sp00-av18} NNOMP.
For this particular observable and energies of our study,
we find very similar results among all the NNOMP i.e.
irrespectively of the reference potential.
Again we find little differences  between the $g$- and $t$-matrix
approaches above $\sim$500 MeV. Thus, medium effects in the interaction
are rather weak (albeit not neglegible) at these higher energies. 
Such is not the case below 400 MeV, particularly in the case 
of Pb, where a clear departure of the $t$-matrix with respect
to the $g$-matrix results is observed. 
However, these differences are  smaller than the ones
due to the presence of the imaginary part of the NNOMP. 
In fact, we can see clear differences within the $g$-matrix
approach by including and suppressing the
imaginary part of the NNOMP at energies above 500 MeV. 
Since  reaction cross section data are scarce above 700 MeV 
(only two data points for all four targets), 
the curves (marked  with triangles) constitute a high energy prediction 
of our work.

\begin{figure}
\scalebox{0.6}{\includegraphics{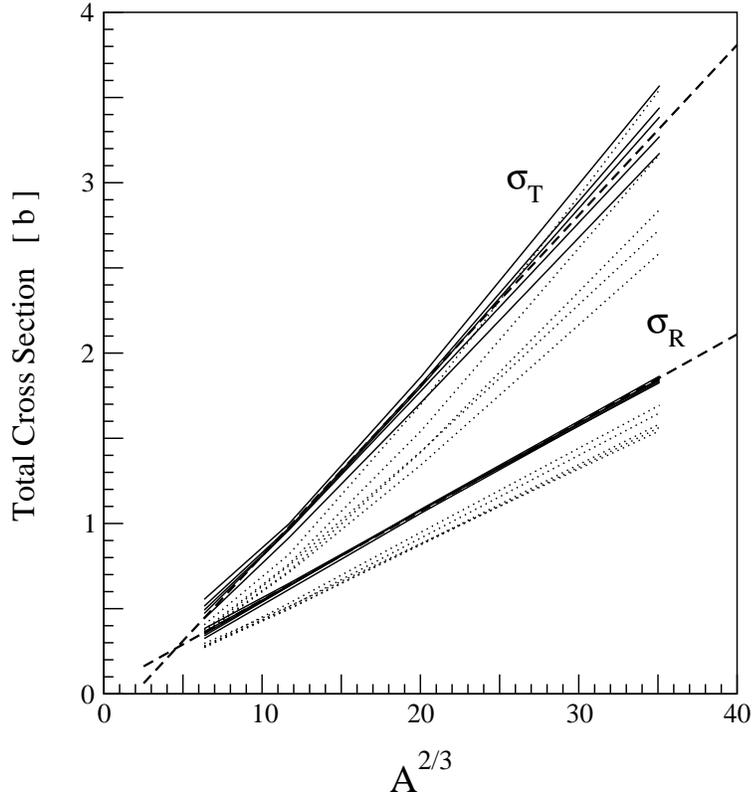}} 
\caption{\label{XA23}
The reaction cross section ($\sigma_R$) and total cross section ($\sigma_T$)
for proton and neutron elastic scattering as function of $A^{2/3}$
for projectile energies of 650, 800, 1040, 1250 and 1500 MeV. 
The solid and dotted curves correspond to $g$-matrix full-folding results 
based on the full and imaginary-suppressed NNOMP respectively.
The dashed curves correspond to straight lines in terms of $A^{2/3}$
(see text for details).
}
\end{figure}

Another feature observed from Figs. \ref{XTotalX4} and \ref{XReactX4}  
is the trend of all cross sections to reach a plateau above 650 MeV. 
We have scrutinized more closely this feature and find that both 
$\sigma_T$ and $\sigma_R$ exhibit an almost linear dependence with $A^{2/3}$, 
with  target masses  16$\leq A\leq$208. 
To illustrate this point we show in Fig. \ref{XA23}
the calculated $\sigma_T$ and $\sigma_R$ as function of 
$A^{2/3}$. The solid curves represent results from $g$-matrix  full-folding
results, based on the {\sc SP00-AV18} full NNOMP, at various energies;
the dotted curves represent the corresponding results without the imaginary  
part of the NNOMP.
The dashed curves serve as reference and correspond to the parametric forms
$\sigma_R =( -0.19 + 0.10 A^{2/3})$ and 
$\sigma_T =( 0.03 + 0.052 A^{2/3})$ in barn units.
Notice, only the full NNOMP results feature a moderate deviation
from the parametric forms shown, which is not the case 
in  calculations where the imaginary part of the NNOMP 
was suppressed (dotted curves).
Thus, the absorptive part of the NNOMP inhibits 
the energy dependence of the total cross sections above 700 MeV.

\subsection{Differential cross sections}

\begin{figure}
\scalebox{0.6}{\includegraphics{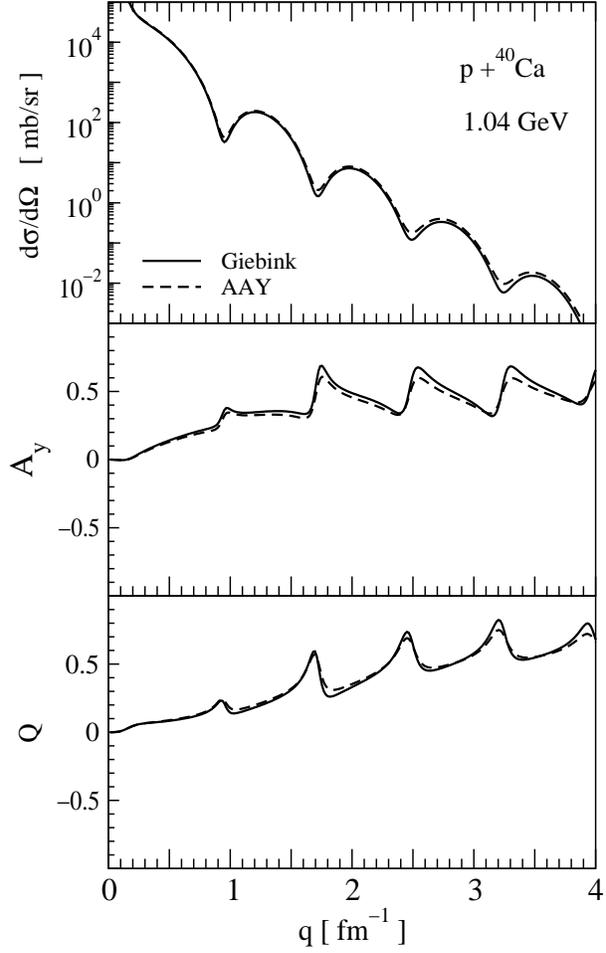}} 
\caption{\label{XAQ_Ca1040_KIN}
         Sensitivity to the kinematics: the calculated differential 
         cross-section (upper frame), analyzing power (middle frame) 
         and spin rotation (lower frame) as functions of the momentum 
         transfer for p+$^{40}$Ca elastic scattering at 1.04 GeV.
         The results correspond to nonrelativistic {\em in-medium} 
         full-folding optical potentials using Giebink's (solid curves)
         and AAY (dashed curves) relativistic kinematics. 
        }
\end{figure}

Differential cross sections and spin observables for elastic 
scattering remain a challenge for any microscopic theory.
An assessment of  alternative relativistic kinematics is made with
a comparison of these observables using Giebink and AAY relativistic 
approaches. 
These results are    illustrated in Fig. \ref{XAQ_Ca1040_KIN}, 
where the differential cross section $d\sigma/d\Omega$, 
analyzing powers $A_y$ and spin rotation functions $Q$ 
for p+$^{40}$Ca elastic scattering at 1.04 GeV are shown as 
function of  momentum transfer $q$.  
We have used $g$-matrices  together with 
Giebink (solid curves) and AAY (dashed curves) kinematics.
This figure shows the case where the differences
should be most pronounced.
Indeed, all the other cases show nearly complete
overlap between the two cases. Nevertheless,
 Fig. \ref{XAQ_Ca1040_KIN} shows that the two prescriptions 
yield  similar results for all observables. Some 
differences can be seen in $A_y$, at momentum transfers above
1.5 fm$^{-1}$, where the AAY  
lie slightly above the  Giebink kinematics results.
We selected
for all further calculations  the   Giebink kinematics.

Next, we study  six NNOMP  in  NA
elastic scattering  at various energies with the full-folding model 
using the $g$-matrix approach. 
We have chosen scattering of protons from $^{40}$Ca and $^{208}$Pb
for which there is a large body of high precision data over a wide energy
range \cite{data}.
The applications used the full NNOMP with reference to
{\sc sp00-nij1}, {\sc -nij2},  {\sc -pari}, {\sc -av18}, {\sc -re93}
and {\sc -invs} solutions.
As in most cases the differences among  these reference NN potentials 
are  moderate, we have chosen  all curves with a single
pattern. 
This also helps to illustrate the level of sensitivity of NA scattering
upon on-shell equivalent potentials with different off-shell behavior.

\begin{figure}
\scalebox{0.7}{\includegraphics{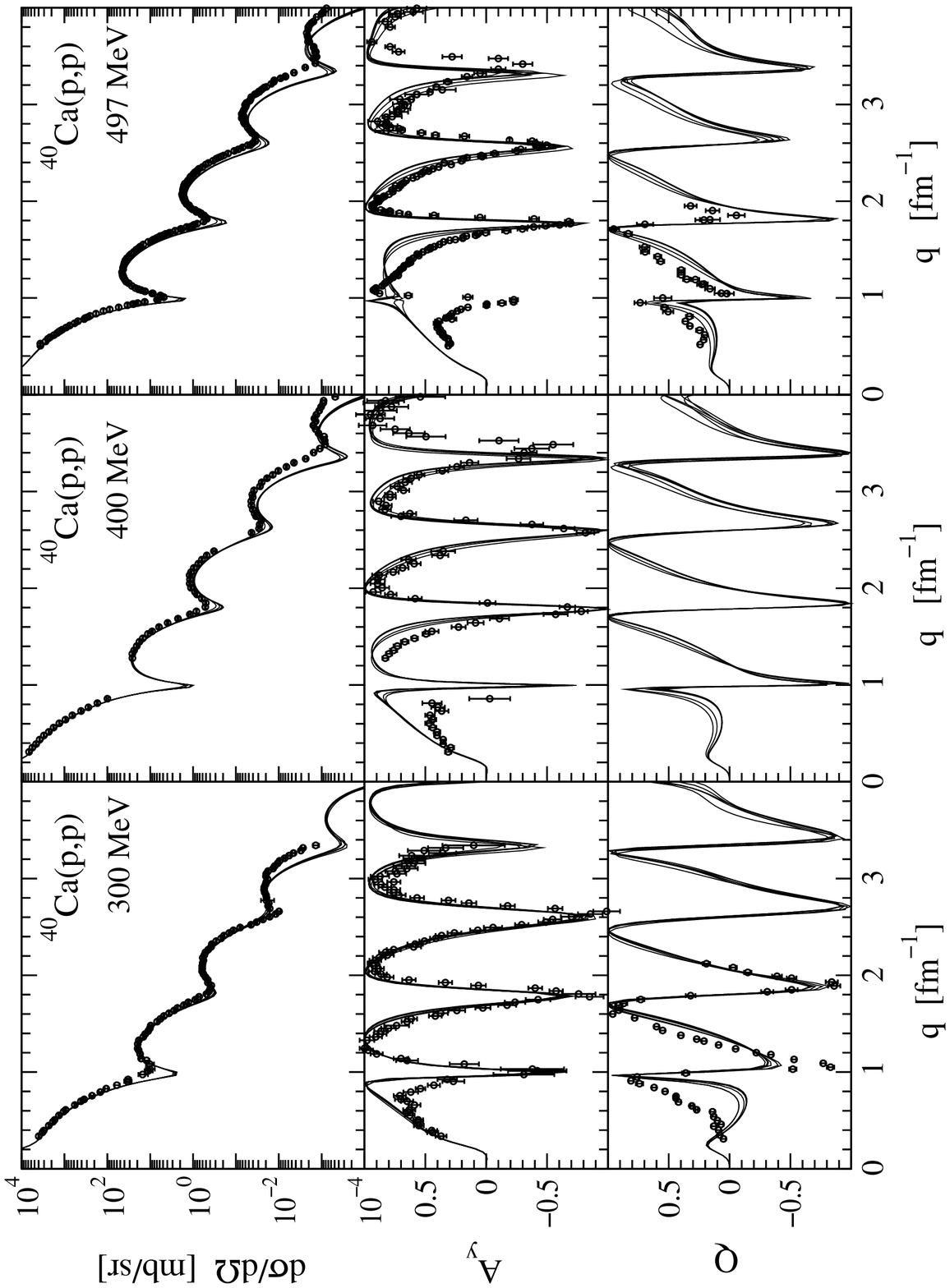}} 
\caption{\label{XAQ_Ca3-5}
         Calculated differential cross-section (upper frames), 
         analyzing power (middle frames) and spin rotation (lower frames) 
         as functions of the momentum transfer for p+$^{40}$Ca 
         elastic scattering at 300\, MeV, 400\,MeV and 497.5\,MeV.
         All curves are obtained from {\em in-medium} full-folding 
         optical potentials using relativistic kinematics. 
         The data are from Ref. \cite{data}.
        }
\end{figure}

\begin{figure}
\scalebox{0.7}{\includegraphics{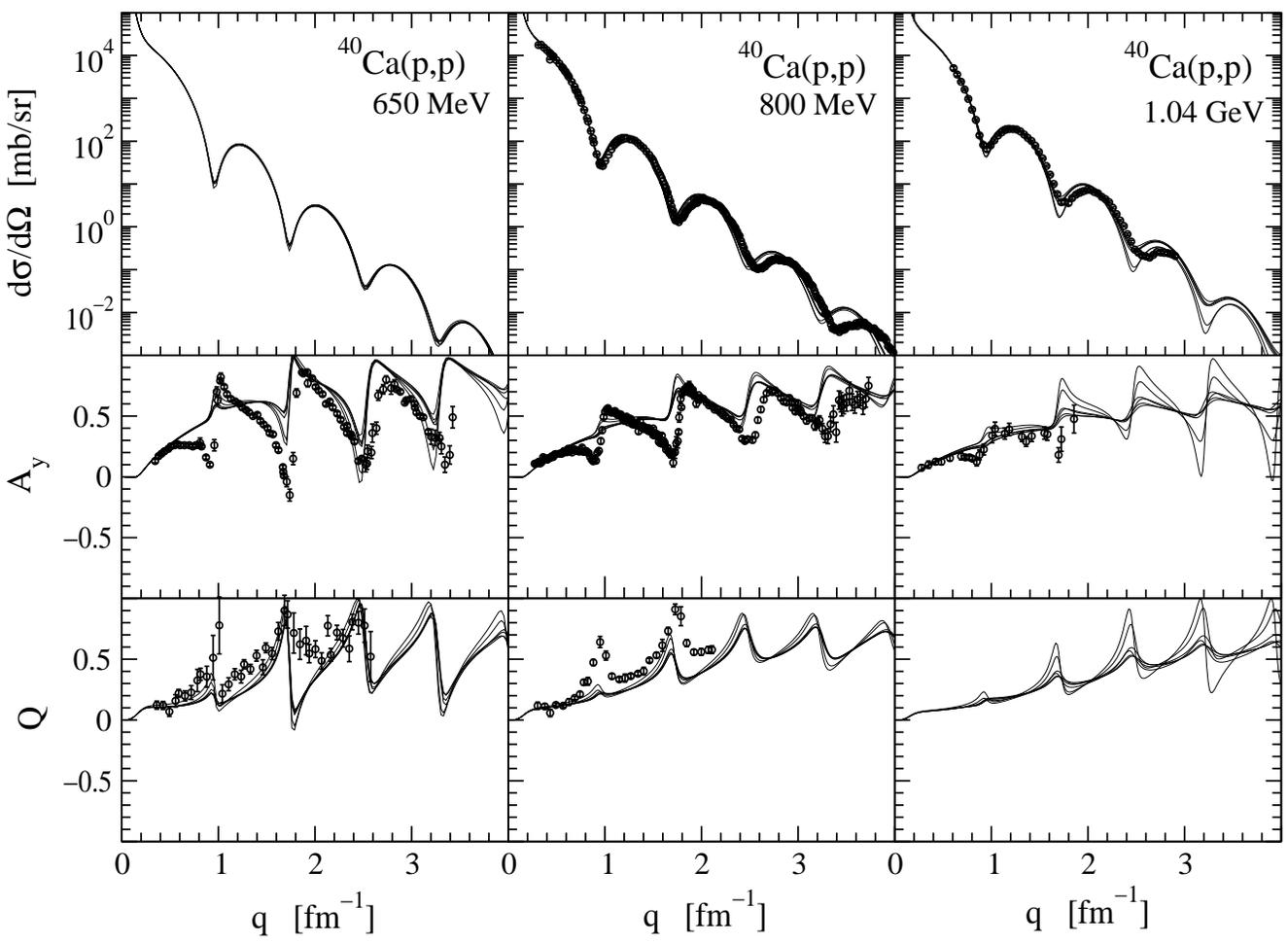}}  
\caption{\label{XAQ_Ca6-10}
         The same as Fig. \ref{XAQ_Ca3-5} but at 650\,MeV, 800\,MeV, and  
         1.04\,GeV.
         The data are from Ref. \cite{data}.
        }
\end{figure}

\begin{figure}
\scalebox{0.7}{\includegraphics{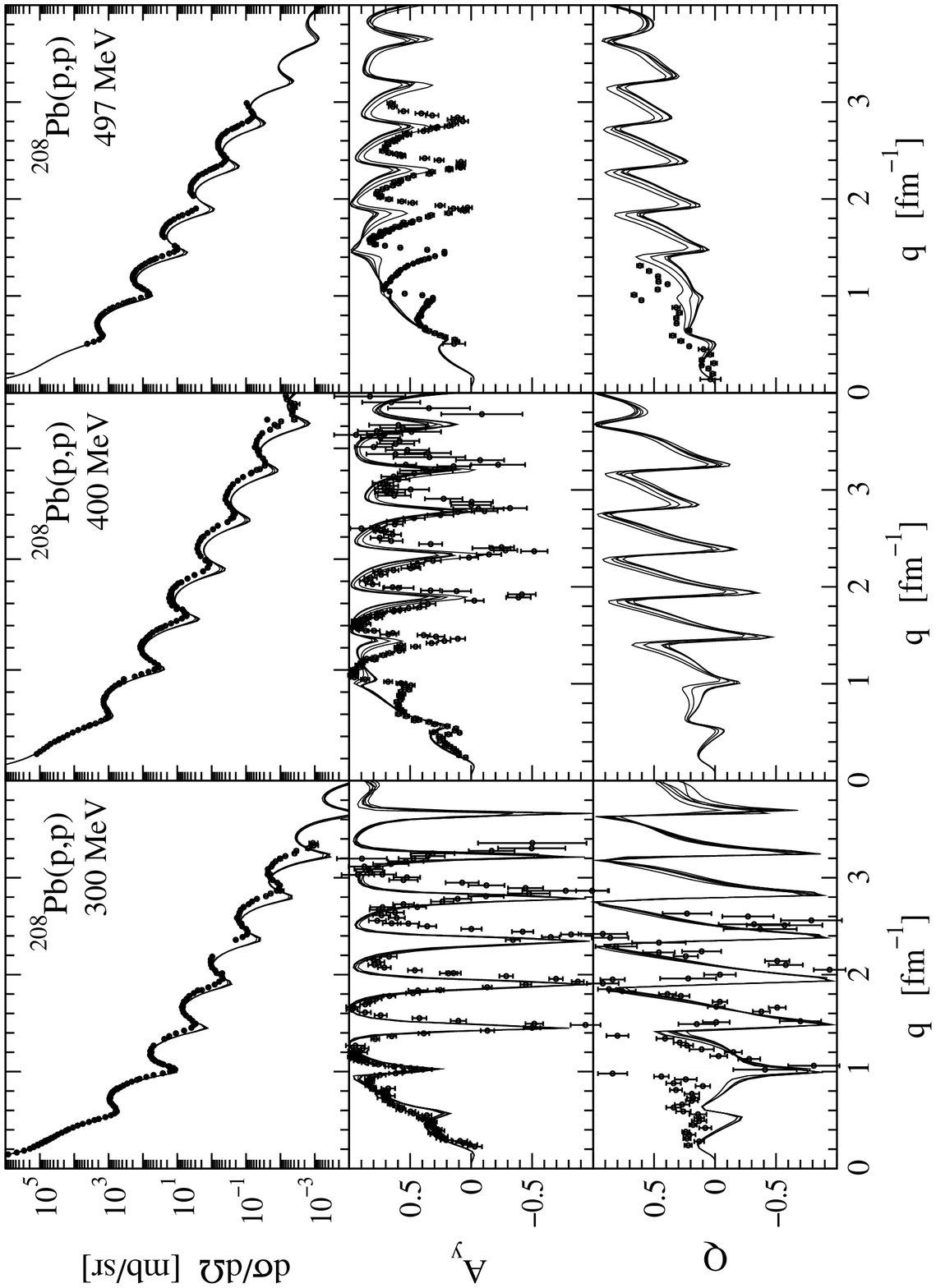}}  
\caption{\label{XAQ_Pb3-5}
         Calculated differential cross-section (upper frame), analyzing power
        (middle frame) and spin rotation (lower frame) 
         as functions of the momentum transfer for p+$^{208}$Pb 
         elastic scattering at 300\, MeV, 400\,MeV and 497.5\,MeV.
         All curves are obtained from {\em in-medium} full-folding 
         optical potentials using relativistic kinematics. 
         The data are from Ref. \cite{data}.
        }
\end{figure}

Considering that the results presented here correspond to
parameter free calculations, the overall description of all elastic 
scattering data is remarkably good. 
Only a close examination of the results shows the 
limitations of our approach.
In Fig. \ref{XAQ_Ca3-5} for p+$^{40}$Ca scattering at 300, 400 and 497.5 MeV, 
for instance, we observe a very good agreement between the full-folding model
results for all observables and  momentum transfers above $\sim$1 fm$^{-1}$.
However, some discrepancies  remain in the description of  low momentum 
transfer data. In Fig. \ref{XAQ_Ca6-10} are shown p+$^{40}$Ca
results  for energies 
between 650 and 1040 MeV. We observe two curves slightly separated
from the rest. Such curves correspond to the {\sc sp00-av18} and 
{\sc sp00-pari} NNOMP, affecting mainly the differential cross section
and analyzing powers at 800 and 1040 MeV.
The possible cause of this feature,
and the fact that this is more pronounced in the case of Ca than in Pb
is not fully understood but numerical reasons cannot be ruled out.
A similar trend is observed in Fig. \ref{XAQ_Pb3-5} for 
p+$^{208}$Pb scattering. 
These limitations have already  been noticed in previous nonrelativistic 
full-folding calculations. Thus,
the identification of a unique  missing element in the approaches, 
to account for the low $q$ failures  remains an open issue.  
Some possibilties have been discussed elsewhere \cite{Ray92}.

\begin{figure}
\scalebox{0.7}{\includegraphics{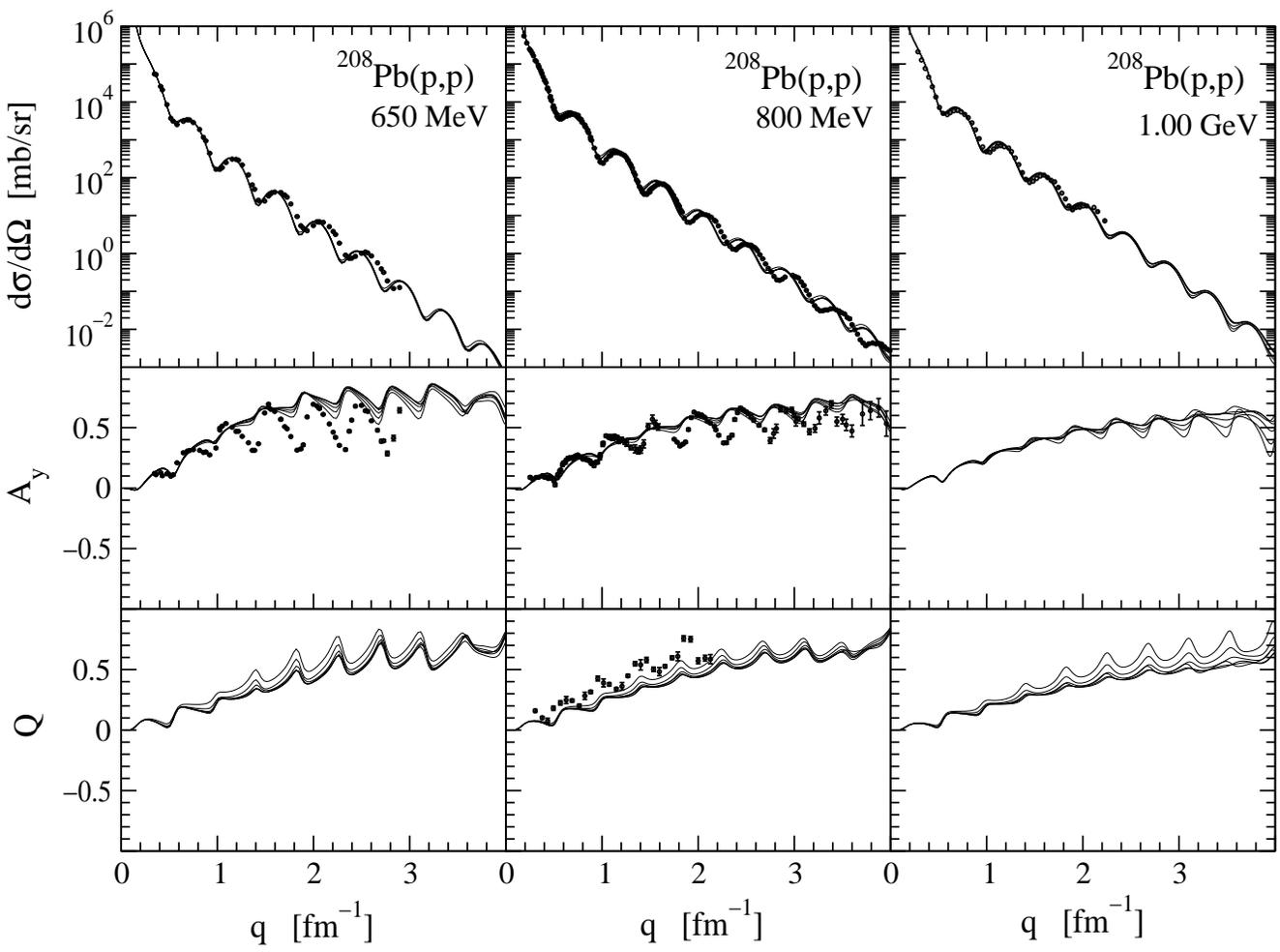}}  
\caption{\label{XAQ_Pb6-10}
         The same as Fig. \ref{XAQ_Pb3-5} but at 650, 800, and 1000 MeV.
         The data are from Ref. \cite{data}.
        }
\end{figure}

In Figs. \ref{XAQ_Pb3-5} and \ref{XAQ_Pb6-10} we show results for 
p+$^{208}$Pb scattering at energies between 300 MeV and 1 GeV. 
With the exception of the 400 and 497.5 MeV cases, the cross section
is very well described by our calculations. 
With respect to the spin observables, there is a
tendency to lose structure  relative to the measured values 
at energies above 650 MeV.

\subsection{Approximations}

\begin{figure}
\scalebox{0.7}{\includegraphics{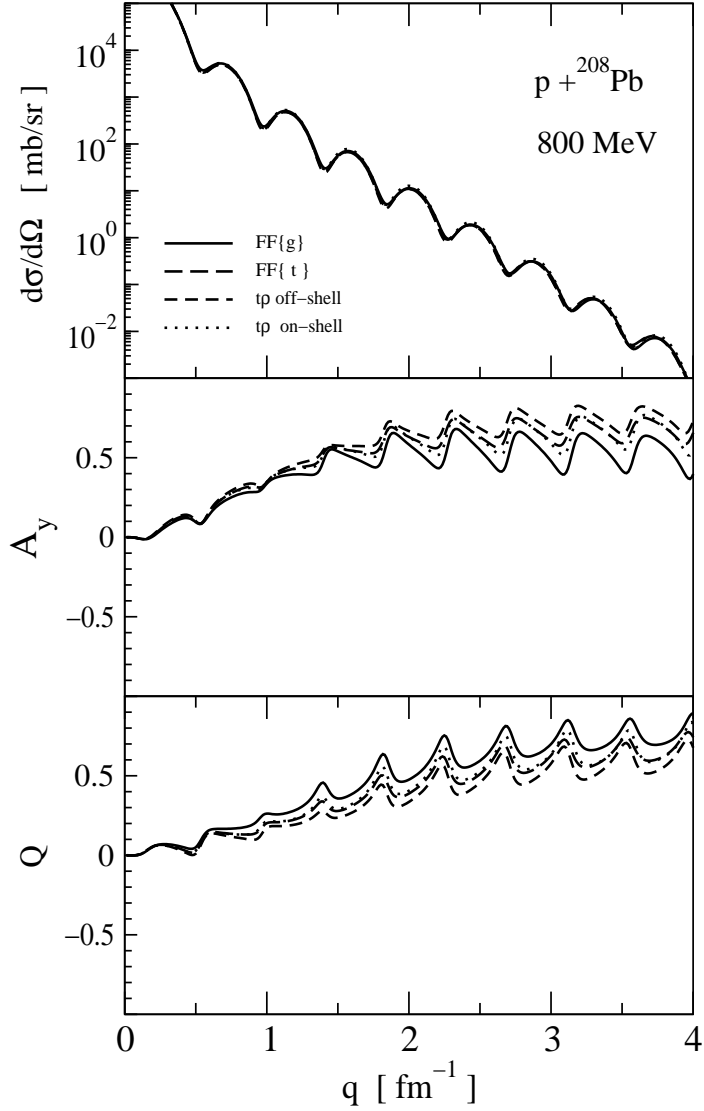}}
\caption{\label{XAQ_FF_TRho}
         A comparison of $g$-matrix full-folding (solid curves), 
         free $t$-matrix full-folding (long-dashed curves)
         off-shell `t$\rho$' (short-dashed curves) and on-shell `t$\rho$' 
         (dotted curves) results
         in the case of p+$^{208}$Pb elastic scattering at 800 MeV.
        }
\end{figure}

The calculation of optical potentials, within the full-folding approach 
used herein, are made without refuge
in assumptions either about the local 
structure of the NN effective interaction or about  the final structure 
of the NA coupling.
In fact, these potentials are treated as non-local operators and are the result of 
a detailed account of the NN effective interaction off-shell. 
In contrast, the optical potentials in the nonrelativistic impulse 
approximation, discussed in Ref. \cite{Ray92}, are assumed local and constructed
with only on-shell $t$-matrix elements as effective interaction.
Important differences have been observed between these two approaches when 
applied to intermediate energy (200-400 MeV) NA scattering.
An illustration of these differences is made in Fig. \ref{XAQ_FF_TRho} for
800 MeV p+$^{208}$Pb elastic scattering, where the differential 
cross section, analyzing powers and spin rotations are shown as 
function of the momentum transfer.
The $g$-matrix full-folding results are represented with solid curves. 
The $t$-matrix full-folding results are represented with long-dashed curves,
whereas short-dashed and dotted curves are used to represent the results
of the off-shell $t\rho$ and on-shell $t\rho$ results respectively.
Clear differences can be seen between the $g$- and $t$-matrix results,
particularly for the spin observables. 
These are pronounced for $q\gtrsim$ 1.5 fm$^{-1}$ which illustrates the   
level of sensitivity to medium effects at these momentum transfers and for 
this particular case.
Within the same $t$-matrix approach however, the full-folding results 
are similar to those obtained within the $t\rho$ approximation. 
The extent of these sensitivities is comparable to contributions from
short range correlations \cite{Ray92}. 

The full-folding calculations presented here are first of their kind
to be tested at energies as high as $\sim$1.5 GeV. 
Differences are observed between these  and previous results,
 particularly in the 800 MeV applications
at forward angles \cite{Ray92}. 
Apart from the locality issue in the NA coupling and the treatment of 
the NN effective interaction off-shell, additional effects have been 
included in those local potential
 calculations and  which may cause variations between the results.
Particularly relevant seem to be
the short range correlations and 
electromagnetic spin-orbit contributions to the NA coupling. 
A careful assessment of these effects in the context of the full-folding  
approach is needed. 
Considering that calculations in the $t\rho$ scheme are much simpler 
than those in the full-folding approach, the former becomes quite
suitable for exploratory purposes.
Quantitative comparisons, however, do require the inclusion of medium 
effects within the full-folding approach.


 \section{Summary and conclusions}
The nonrelativistic full-folding optical model approach for
NA scattering has been extented  into the relativistic regime.
Kinematical issues involving the
off-shell Lorentz boost of the colliding particles between
the NN and the NA c.m. reference frames have been addressed. Also,
explicit nuclear medium effects have been  incorporated with the use of
microscopic NN effective interaction as obtained in the framework of 
nuclear matter $g$-matrix using an NNOMP which fully accounts 
for the inelasticities and isobar resonances at nucleon energies as
high as $\sim$3\,GeV.
The nuclear matter $g$-matrices were obtained considering both
Pauli blocking and  selfconsistent nuclear fields as
in the traditional Brueckner theory. 
Minimal relativity corrections were extracted from the 
Brockmann and Machleidt approach to relativistic nuclear matter.
Effects arising from Wigner rotations and electromagnetic spin-orbit 
corrections were not included. 

The study considered both $t$-matrix and {\em in-medium}
self-consistent $g$-matrix approaches. 
With the inclusion of relativistic kinematics corrections,
in conjunction with a realistic description of NN resonances 
and inelasticities by means of an NNOMP, we obtain a  good 
description of both the total and the differential scattering 
observables for NA elastic scattering.
The results exhibit a weak sensitivity to the choice of the relativistic
approach --AAY or Giebink-- to correct the kinematics.
We also observe that medium effects are significant over the whole 
energy range of our study, although they are rather weak above 400 MeV. 
In contrast, the inelasticities of the NN interaction become important 
above 400 MeV as was observed particularly in 
the description of the total cross sections. 

Although our study allows a reasonable description of the differential
observables at energies as high as 1 GeV in NA scattering, 
specific details remain to be explained; notably
the low $q$ behavior of the spin observables in the 400-500 MeV range.
As our primary effort has been to provide a parameter free 
nonrelativistic framework for the study of NA elastic scattering 
with a minimal account of relativistic effects at these high energies,
a systematic study of various other effects has not  been pursued
in order to keep our discussion focused.
However, future work will
scrutinize the systematics of the 
calculated observables under the use of alternative densities, mixed 
density representations, electromagnetic effects 
and higher order correlations.
Off-shell effects arising from the non-locality range of the separable
description of the NNOMP above pion threshold also require further
investigation. Nevertheless,  the level of agreement we have
achieved, within the nonrelativistic approach, is comparable to
what has been obtained within relativistic approaches.

The study presented  is  limited in its value as we have not used 
a covariant  two-body and (A+1)-body dynamics. 
The approach adopted is a practical one and is largely motivated and justified by 
the good agreement of the  numerical results with data. 
Thus we  claim, that use of minimal  relativity 
in conjuction with  NN optical model interactions, which fully account for the 
inelasticities and isobar resonances above the pion threshold,
yield quantiative descriptions of   NA scattering 
up to energies of 1.5 GeV.

 \appendix

 \section{Relativistic kinematics}
%
%

We have  used two prescriptions for the relativistic kinematics 
involved in the transformation of the two-body colliding 
momenta between the pair c.m. and the projectile-nucleus c.m.
These schemes have been developed by 
Giebink \cite{Gie82} and by 
Aaron {\em et al.} \cite{Aar68}.                 
The latter has also been obtained by imposing time reversal
invariance on the scattering amplitude \cite{Ern80}.

Following Giebink in the context of a manifestly 
Lorentz invariant two-body transition amplitude, let 
$k=(\bar\omega,{\bf k})$ and $p=(\bar\varepsilon,{\bf p})$
be the projectile 
and struck-nucleon four-momenta in the projectile-nucleus c.m.
The corresponding momenta in the exit channel are represented with
$k^\prime$ and $p^\prime$. 
In Giebink's approach the total four-momentum is conserved,
\begin{equation}
\label{conserv}
k+p = k^\prime + p^\prime \equiv Q \ ,
\end{equation}
which, to be valid, requires each particle to be off mass-shell.
The relative momenta in the two-body c.m. is readily obtained by 
applying a Lorentz transformation to  momenta with a single
boost $\beta$. 
This boost is obtained from  the invariant
\begin{equation}
\label{S}
(\bar\omega + \bar\varepsilon)^2 - ({\bf k} + {\bf p})^2 = S \ ,
\end{equation}
and from  
\begin{equation}
\label{beta_gie}
{\mathbf{\beta}} = 
\frac{{\bf k} + {\bf p}}{\bar\omega + \bar\varepsilon} =
\frac{{\bf k}^\prime+{\bf p}^\prime}{\bar\omega^\prime+\bar\varepsilon^\prime} 
\  .
\end{equation}
With this velocity, a direct calculation yields 
for the incident (${\bf k_r}$) and 
outgoing (${\bf k_r^\prime}$) relative momenta
\begin{equation}
\label{k_gie}
{\bf k_r} = \frac{ \left 
               ( \bar\varepsilon  + \varepsilon_r \right ) {\bf k} - 
                 \left ( \bar\omega + \omega_r \right ) {\bf p}}
               {\bar\varepsilon + \varepsilon_r + \bar\omega +
               \omega_r } \ ,
\quad
{\bf k_r}^\prime = \frac{ \left ( \bar\varepsilon^\prime  + 
                     \varepsilon_r^\prime \right ) {\bf k}^\prime - 
                    \left ( \bar\omega^\prime + 
                            \omega_r^\prime \right ) {\bf  p}^\prime}
               {\bar\varepsilon^\prime + \varepsilon_r^\prime + 
                \bar\omega^\prime + \omega_r^\prime } \ .
\end{equation}
In these expressions the subscript `r' denotes the on mass-shell 
relative energy
\begin{equation}
\label{rel_ene}
\omega_r = \sqrt{m_p^2 + {\bf k_r}^2 } \ ,
\quad
\varepsilon_r = \sqrt{m_t^2 + {\bf k_r}^2 } \ ,
\end{equation}
where $m_p$ and $m_t$ represent the masses of the projectile and 
struck-nucleon respectively.
In all the above expressions the magnitude of the relative momentum 
${\bf k_r}^2$ is needed. 
It can be shown that
\begin{equation}
\label{k2_gie}
{\bf k_r}^2 = \frac{1}{4S}\xi^2(S,k^2,p^2) \ ,
\end{equation}
where the $\xi$-function is defined as
\begin{equation}
\label{xi}
\xi(x,y,z) = \sqrt{ (x-y-z)^2 - 4yz}\ .
\end{equation}
Notice that the $\xi$-function in Eq. (\ref{k2_gie}) is evaluated at 
the off-mass-shell invariants $k^2= \bar\omega^2 - {\bf k}^2$
and $p^2= \bar\varepsilon^2 - {\bf p}^2$.
The actual implementation of Giebink's procedure faces the difficulty of  
improper Lorentz transformations which occur when $S$ in Eq. (\ref{S}) 
becomes negative; this happens for very large momenta ${\bf k}+{\bf p}$.
However, as the bound-state wavefunctions of the struck nucleons confine
the momentum distribution of ${\bf p}$ to magnitudes below $\sim$2 fm$^{-1}$, 
such improper contributions occur for very large ${\bf k}$ and ${\bf k}'$, 
affecting only far off-shell matrix elements in $U({\bf k}',{\bf k})$.
A way to circumvent this difficulty is to restrict $S$ in Eq. (\ref{S})
near on-mass-shell. 
Thus, we approximate
\begin{equation}
\label{S0}
S \approx m_p^2+m_t^2+2\bar\omega\bar\varepsilon-2{\bf k}\cdot{\bf p}\  .
\end{equation}
Averaging the Fermi motion of the target nucleons (${\bf p}$) 
allows the simplification
\begin{equation}
\label{S0app}
S \rightarrow S_\circ = m_p^2+m_t^2+2\bar\omega\bar\varepsilon \ .
\end{equation}
Consistent with the above, the following 
forms are obtained for the relative energies
\begin{equation}
\omega_r = \frac{\bar\omega \bar\varepsilon + m_t^2}{\sqrt{S_\circ}} ,
\quad\mbox{and}\quad
\varepsilon_r = \frac{\bar\omega \bar\varepsilon + m_p^2}{\sqrt{S_\circ}} \ .
\end{equation}
Clearly $\sqrt{S_\circ} = \omega_r + \varepsilon_r$,
which is  a handy result
for the energy denominators in Eq. (\ref{k_gie}).
The full-folding calculations following Giebink relativistic 
kinematics were made using $\bar \varepsilon=\bar \varepsilon'=M$
and $\bar \omega=\bar \omega'=M+E_{lab}$, with $M$ the nucleon mass.  

%
%

In the relativistic prescription developed by Aaron and 
collaborators \cite{Aar68}, the relative momenta for the incoming 
and outgoing channels require different boost velocities.
In each channel the particles are set on-mass-shell and
the corresponding boost is represented by Eq. (\ref{beta_gie}). 
The resulting relative momentum exhibits the same structure as the
one given by Eq. (\ref{k_gie}) but with the substitution 
$\bar\omega\rightarrow\omega({\bf k})$ and
$\bar\varepsilon\rightarrow\varepsilon({\bf p})$. 
A direct calculation yields for the magnitude
\begin{equation}
\label{k_aay}
{\bf k}_r^2 = \frac{1}{4s_{in}}\xi^2(s_{in},m_p^2,m_t^2) \ ,
\end{equation}
where 
\begin{equation}
\label{s}
s_{in} = (\varepsilon({\bf p})+\omega({\bf k}))^2 - ({\bf p} + {\bf k})^2 \ .
\end{equation}
An analogous result is obtained for the outgoing channel.

 \section{Algorithm to determine the NNOMP from data}
%
%

Consider that there are three distinct  Hamiltonians \cite{Fun01}. 
They are the  {\em reference}  Hamiltonian  $H_0$, 
a {\em projected} Hamiltonian $H_{PP}$,
and a {\em full optical model} Hamiltonian $\cal H$. The first of these, the 
reference Hamiltonian $H_0:=T+V_0$, invokes a given potential $V_0$ 
for which one can find Schr\"odinger equation reference  
solutions. The physical outgoing solutions
$\psi_0:=\psi_0^+({\bf r,k},E)$ of $H_0$ we suppose yields a 
unitary  S-matrix. 
We assume further that this Hamiltonian is completely 
specified such that
evaluation of any quantity, wave function, S-matrix, 
K-matrix {\em etc.} is facilitated. The Feshbach projection 
operator formalism  gives 
the projected  Hamiltonian, $P H_0P=H_{PP}$.  
We presuppose 
completeness, $P+Q=1$, 
and a finite rank representation of the Q space
\begin {equation} \label{eqn_III.17}
Q:=\sum_{i=1}^N |\Phi_i><\Phi_i|=\sum_{i=1}^N |i><i|,
\end{equation}
with the Q space basis functions $|\Phi_i>$ interpreted 
as doorway states. With these doorway states we make the link 
between  the  QCD and  the hadronic sectors; the latter  
encompassing
nucleons, mesons and other free particles.
Thus we   assume that meson creation/annihilation 
occurs only in the highly nonlinear 
QCD sector so that   Q space wave functions are  projections
of such  processes onto hadronic particle coordinates. 
The third of our
Hamiltonians, the {\em full optical model} Hamiltonian,
comprises the reference Hamiltonian
$H_0$ and the proper optical model potential $\cal V$. 
That potential is  complex and nonlocal, and taken to be
 separable of finite rank,  
${\cal H}:=T+V_0+{\cal V}(r,r';lsj,E)$. 

The Schr\"odinger equation specified with $\cal H$ 
has regular physical solutions  
$\Psi^+:=\Psi^+({\bf  r, k},E)$ whose  asymptotic boundary conditions 
we deem to match with the {\em experimental} elastic channel S-matrix. 
Specifically, for these experimental S-matrices we have used  the continuous  
solutions SP00 from  VPI/GWU \cite{SAID}.
The reference potential $V_0$ and separable potential
form factors are still to be specified in detail 
with the application.

To obtain the optical potential on the basis 
of a given reference potential,
we express first the solutions of the projected 
Hamiltonian
in terms of the reference Hamiltonian and 
the {\em a priori} defined 
Q space projector. The Lippmann--Schwinger equation, 
\begin{equation} \label{eqn_III.19}
|\psi_P>=|\psi_0>-\sum_jG^+_0|j><j|H_{QP}|\psi_P>,
\end{equation}
is still very general and does not 
depend upon a specific
representation. However, in the following we assume 
a partial wave expansion and the following equations are identified as radial equations
with the set of quantum numbers suppressed.    

As projector orthogonality $PQ=QP=0$ implies that 
\begin{equation}\label{eqn_III.21}
<j|H_{QP}|\psi_P>=\sum_i^N 
\{ <\Phi|G^+_0|\Phi>\}^{-1}_{ji}<i|\psi_0>,
\end{equation}
the  solutions of Eq.\,(\ref{eqn_III.19})  
can be written in terms of $|\psi_0>$ as 
\begin{equation}\label{eqn_III.22}
|\psi_P>\ =\ |\psi_0> -\sum_{ij}^N G^+_0|i>
\{ <\Phi|G^+_0|\Phi>\}^{-1}_{ij}<j|\psi_0>\
=\ |\psi_0>-\sum_{ij}^N G^+_0 \Lambda_{ij}|\psi_0>,
\end{equation}
wherein one can identify a separable potential
\begin{equation}\label{eqn_III.23}
|i>\{<\Phi|G^+_0|\Phi>\}^{-1}_{ij}<j|\ =\ |i>
\lambda_{ij}<j|=:\Lambda_{ij}(r,r').
\end{equation}
Note then that definition of  Q space gives  a 
 specification of the separable  
strengths $\lambda_{ij}(lsj,E)$ that is unique.
 The resultant Eq.\,(\ref{eqn_III.22}) has the 
form of a first order Born approximation but in 
fact it is an  exact result.

To proceed,  initially we abandon the exactitude  of
Eq.\,(\ref{eqn_III.22}) and require the strength matrix, 
\begin{equation}\label{eqn_III.24}
\lambda_{ij}\ =\ \{<\Phi|G^+_0|\Phi>\}^{-1}_{ij},
\end{equation}
to be  constrained asymptotically by the experimental
 S-matrix of the full Hamiltonian
Schr\"odinger equation, {\em i.e.} asymptotically we 
induce  $|\psi_P>=|\Psi_{\cal H}>$.
This implies that  complex optical model
strengths $\lambda_{ij}$ emerge 
as a result of matching to Riccati-Hankel functions 
and non unitary S-matrices with 
\begin{equation}\label{eqn_III.25}
|\Psi_{\cal H}>= |\psi_P>\sim \frac{1}{2i}\left(-h^-(rk) 
+h^+(rk) S(k)\right).
\end{equation}
The strengths $\lambda_{ij}$ then can be simply determined from the 
linear system of equations 
\begin{equation}\label{eqn_III.26}
\frac 1{2i}{h^+(Rk)\left(S(k)-S_0(k)\right)= 
\sum_{ij}G^+_0|i>\lambda_{ij}<j|\psi_0^+>}.
\end{equation}
To reinforce a Lippmann--Schwinger equation, with 
the experimental S-matrix
as  boundary condition or equivalently with 
strengths $\lambda_{ij}$ from Eq.\,(\ref{eqn_III.26}), 
a transformation  of the separable potential 
Eq.\,(\ref{eqn_III.23}) is made.
This is achieved with 
\begin{equation} \label{eqn_III.27}
{\cal V}(r,r'):=\Lambda\frac{1}{(1-G^+_0\Lambda)},
\end{equation}
which contains the separable potentials as defined with
Eq.\,(\ref{eqn_III.23}) but whose strengths 
now are solutions of Eq.\,(\ref{eqn_III.26}).  
As the transformation Eq.\,(\ref{eqn_III.27}) 
contains  integration of orthonormal functions,
 only strengths are altered. Using this optical model in  
the full Hamiltonian, physical solutions are obtained with 
reference solutions $|\psi_0>$ and   Greens function $G^+_0$ of the
reference Hamiltonian $H_0$ by means of the 
Lippmann--Schwinger equation
\begin{equation}\label{eqn_III.28}
|\Psi_{\cal H}>=|\psi_0>+G^+_0{\cal V}|\Psi_{\cal H}>.
\end{equation}

\begin{acknowledgments}
H. F. A  acknowledges partial support from FONDECYT under grant 1970508.
The authors thank Profs. B. C. Clark and L. Ray for making 
available much of the differential scattering data presented in this work.
\end{acknowledgments}


 \newpage

%

 \end{document}